\documentclass{aa}
\usepackage{natbib}
\usepackage{graphicx}

\title{The AKARI/IRC Mid-Infrared All-Sky Survey}
\author{
Daisuke Ishihara$^{1,2}$,
Takashi Onaka$^2$,
Hirokazu Kataza$^3$,
Alberto Salama$^4$,
Carlos Alfageme$^4$
\thanks{Present address: INTA, Ctra. de Ajalvir km. 4. 28850 Torrej\'{o}n de Ardoz, Madrid},
Angelo Cassatella$^{4,5,6}$,
Nick Cox$^4$
\thanks{Present address: Instituut voor Sterrenkunde, Katholieke
Universiteit Leuven, Celestijnenlaan 200D, 3001 Leuven, Belgium},
Pedro Garc\'{i}a-Lario$^4$,
Craig Stephenson$^4$
\thanks{Present address: Deimos Space S.L.. Ronda de Poniente, 
19 Edificio Fiteni VI, 28760 Tres Cantos. Madrid},
Martin Cohen$^7$,
Naofumi Fujishiro$^{3,8}$
\thanks{Present address: Cybernet system Co. Ltd., 3
Kanda-neribeicho, Chiyoda-ku, Tokyo, Japan, 101-0022},
Hideaki Fujiwara$^2$,
Sunao Hasegawa$^3$,
Yoshifusa Ita$^9$,
Woojung Kim$^{3,2}$
\thanks{Present address: SONY Co. Ltd.,
4-14-1, Asahi-cho, Atsugi-shi, Kanagawa, Japan, 243-0014},
Hideo Matsuhara$^3$,
Hiroshi Murakami$^3$,
Thomas G. M\"uller$^{10}$,
Takao Nakagawa$^3$,
Youichi Ohyama$^{11}$,
Shinki Oyabu$^3$,
Jeonghyun Pyo$^{12}$,
Itsuki Sakon$^2$,
Hiroshi Shibai$^{13}$,
Satoshi Takita$^3$,
Toshihiko Tanabe$^{14}$,
Kazunori Uemizu$^3$,
Munetaka Ueno$^3$,
Fumihiko Usui$^3$,
Takehiko Wada$^3$,
Hidenori Watarai$^{15}$,
Issei Yamamura$^3$,
and Chisato Yamauchi$^3$
}
\institute{
Department of Physics,
Nagoya University, Furo-cho, Chikusa-ku, Nagoya, Aichi, 464-860, Japan
\and
Department of Astronomy, Faculty of Science,
University of Tokyo, 3-1-1 Hongo, Bunkyo-ku, Tokyo, 113-0003, Japan
\and
Institute of Space and Astronautical Science (ISAS),
Japan Aerospace Exploration Agency (JAXA), 3-1-1, Yoshinodai,
Sagamihara, Kanagawa, 229-8510, Japan
\and
European Space Astronomy Center (ESAC),
Villanueva de la Ca\~{n}ada, Apartado 78, 28691 Madrid, Sapin
\and
INAF, Istituto di Fisica dello
Spazio Interplanetario, Via del Fosso del Cavaliere 100, 00133 Roma,
Italy
\and
Dipartimento di Fisica, Universita' Roma Tre, Via della Vasca Navale
100, 00146 Roma, Italy
\and
Radio Astronomy Laboratory, University of California, Berkeley, USA,
\and
Department of Physics, Faculty of Science,
University of Tokyo, 3-1-1 Hongo, Bunkyo-ku, Tokyo, 113-0003, Japan
\and
National Astronomical Observatory of Japan, Mitaka, Tokyo, 181-8588, Japan,
\and
Max-Planck-Institut f\"ur extraterrestrische Physik, 
Giessenbachstra$\beta$e, 85748 Garching, Germany
\and
Academia Sinica, Institute of Astronomy and Astrophysics (ASIAA), Taipei 10617, Taiwan
\and
Korea Astronomy and Space Science Institute (KASI), 
61-1, Hwaam-dong, Yuseong-gu, Daejeon, 305-348, Republic of Korea
\and
Graduate School of Science,
Osaka University, 1-1, Machikaneyama, Toyonaka, Osaka, 560-0043, Japan
\and
Institute of Astronomy, Faculty of Sicence, University of Tokyo,
Mitaka, Tokyo, 181-8588, Japan
\and
Space Applications Mission Directorate, Japan Aerospace Exploration
Agency (JAXA), 2-1-1, Sengen, Tsukuba, Ibaraki, 305-8505, Japan
}

\abstract
{AKARI is the first Japanese astronomical satellite
dedicated to infrared astronomy.
One of the main purposes of AKARI is the all-sky survey
performed
with six infrared bands between 9 and 200\,$\mu$m
during the period from 2006 May 6 to 2007 August 28.
In this paper,
we present the mid-infrared part
(9\,$\mu$m and 18\,$\mu$m bands)
of the survey carried out
with one of the on-board instruments, the Infrared Camera (IRC).
}
{We present unprecedented observational results
of the 9 and 18\,$\mu$m AKARI all-sky survey
and detail the operation and data processing leading to the
point source detection and measurements.}
{
The raw data are processed to produce small images for every scan
and point sources candidates, above the 5$\sigma$noise level
per single scan, are derived.
The celestial coordinates and fluxes of the events are determined
statistically and
the reliability of their detections is secured
through multiple detections of the same source
within milli-seconds, hours, and months from each other.
}
{
The sky coverage is more than 90\% for both bands.
A total of 877,091 sources (851,189 for 9\,$\mu$m,
195,893 for 18\,$\mu$m) are confirmed and
included in the current release of the point source catalogue.
The detection limit for point sources is 50\,mJy and 90\,mJy
for the 9\,$\mu$m and 18\,$\mu$m bands, respectively.
The position accuracy is estimated to be better than 2$''$.
Uncertainties in the in-flight absolute flux calibration are 
estimated to be 3\%
for the 9\,$\mu$m band and
4\% for the 18\,$\mu$m band.
The coordinates and fluxes of detected sources in this survey are
also compared with those of the IRAS survey
and found to be statistically consistent.}
{}
\authorrunning{Ishihara, D. et al.}
\titlerunning{AKARI MIR All-Sky Survey}

\begin{document}
\maketitle

\section{Introduction}
Unbiased and sensitive all-sky surveys at infrared wavelengths
are important for the various fields of astronomy.
The first extensive survey of the mid- to far- infrared sky was made
by the Infrared Astronomy Satellite (IRAS) mission
launched in 1983 \citep{IRAS}.
IRAS surveyed 87\% of the sky in four photometric bands
at 12, 25, 60 and 100\,$\mu$m 
and substantially pioneered the various new fields of astronomy,
such as
circumstellar debris disks around Vega-like stars \citep{Vega},
and a new class of galaxies that radiate most of their
energy in the infrared \citep{LIRG}.

A decade later than IRAS, the Midcourse Space Experiment \citep[MSX;][]{MSX}
surveyed the Galactic plane as well as
the regions not observed by or confused in the IRAS mission
with higher sensitivity and higher spatial resolution (18.3$''$)
in four mid-infrared broad bands centered at 8.28, 12.13, 14.65 and 21.23\,$\mu$m
and two narrow bands at 4.29 and 4.35\,$\mu$m.
The MSX catalogue (version 1.2) of the Galactic plane survey contains
323,052 sources, 3 times as many as IRAS listed for the same region.

In 2006, {\it AKARI}, the first Japanese space mission
dedicated to infrared astronomical observations \citep{Murakami}, was
launched and brought into a sun-synchronous polar orbit at an altitude of 700\,km.
It has two scientific instruments, the Infrared Camera 
\citep[IRC;][]{IRC} for 2--26\,$\mu$m
and the Far-Infrared Surveyor \citep[FIS; ][]{FIS} for 50--200\,$\mu$m.
{\it AKARI} has a Ritchey-Chretien type cooled telescope 
with a primary-mirror aperture size of 685\,mm \citep{Tel},
which is operated at 6\,K by liquid helium and mechanical coolers.
One of the major observational objectives of {\it AKARI} is an all-sky survey observation.
The survey was executed
during the life time of the cooling medium
between 2006 May 8 and 2007 August 28.
The 9 and 18\,$\mu$m bands of the IRC
and the 60, 90, 140, and 160\,$\mu$m
bands of the FIS were used for the all-sky survey.

In this paper, we present 
the mid-infrared part of the all-sky survey performed with the IRC.
The IRC was originally designed for imaging and spectroscopic observations
in the pointing mode, however,
the all-sky observation mode was added as an operation mode
following ground tests, 
in which the acceptable performance of continuous survey-type
observations was confirmed
\citep{ScanOpe}.
The data of the IRC all-sky survey observation have
been processed by a dedicated program and a point source catalog
has been prepared.
The content of this paper is based on the $\beta$-1
version of the {\it AKARI}/IRC All-Sky Survey Point Source Catalogue.

The outline of the observation is presented in \S 2.
The data reduction is described in \S 3.
The quality of the catalogue is statistically evaluated in \S 4,
and a summary is given in \S 5.

\section{Observations}

\subsection{The AKARI satellite \label{akari}}
{\it AKARI} has two observational modes: the all-sky survey and pointed observations.
In the pointed observations,
the telescope stares at the target or makes round trip scans
around the target for about 10 minutes.
In the all-sky survey,
the spacecraft spins around the Sun-pointed axis once every orbit
keeping the telescope toward a great circle and
making continuous scans of the sky at a scan rate of 216$''$ s$^{-1}$.
The orbit rotates around the axis of the Earth
at the rate of the yearly round of the earth.
Thus, the whole sky is in principle covered in half a year.

In the 1$^{\rm st}$ half year (Phase 1) of the mission
the all-sky survey dedicated as a first priority
with pointed observations toward
the North Ecliptic Pole and the Large Magellanic Cloud.
In the 2$^{\rm nd}$ and 3$^{\rm rd}$ half year (Phase 2a and 2b),
the all-sky survey was continued until exhaust of the cooling medium (liquid helium).
The time was devided between 
pointed observations
survey observations
to increase the final sky coverage of the all-sky survey.

\subsection{The Infrared Camera (IRC)}
The mid-infrared component of the {\it AKARI} All-Sky Survey was performed
with one of the two focal-plane instruments: the Infrared Camera (IRC).
IRC covers the wavelength range 2--26\,$\mu$m with three independent channels:
NIR (2--5.5\,$\mu$m), MIR-S (6--12\,$\mu$m) and MIR-L (12--26\,$\mu$m).
IRC was primarily designed for deep imaging and spectroscopy in pointed observations.
All the channels have filter wheels which hold
3 filters and 2 spectroscopic dispersers. 
Each channel has a large format array that provides a
wide field-of-view (FOV) 
of $10\arcmin \times 10\arcmin$.
The MIR-S and MIR-L channels have
infrared sensor arrays of $256\times256$ pixels
(Si:As/CRC-744 manufactured by Raytheon).
The pixel scales for MIR-S and MIR-L are
$2\farcs34\times2\farcs34$ and $2\farcs51\times2\farcs39$, respectively.
%
The field-of-views of MIR-S and MIR-L are separated
by 20$'$ in the cross-scan direction.
More details and in-flight performance are described in \citet{IRC}.

\subsection{All-Sky Survey operations of the IRC\label{sec:scanope}}

\begin{table}
\caption{Parameters for Mid-Infrared All-Sky Survey operation.\label{tbl:ircscan}}
\begin{tabular}{lcc}
\hline\hline
Filter band (Camera)& S9W (MIR-S)   & L18W (MIR-L)\\
\hline
Wavelength$^{\ddagger}$  & 6.7--11.6\,$\mu$m & 13.9--25.6\,$\mu$m \\
Isophotal wavelength & 8.61\,$\mu$m & 18.39\,$\mu$m\\
Effective bandwidth          & 4.10\,$\mu$m & 9.97\,$\mu$m \\
Sampling rate (period) & \multicolumn{2}{c}{22.27\,Hz (44\,ms)}\\
Scan rate (exposure$^*$) & \multicolumn{2}{c}{216$''
 $s$^{-1}$ (11\,ms)} \\
Reset rate (period) & \multicolumn{2}{c}{0.074\,Hz (13.464\,s)}\\
Operation           & \multicolumn{2}{c}{$256\times2$\, pix}\\
Operated row &  \multicolumn{2}{c}{$117^{th}$, $125^{th}$}\\
Binning             & \multicolumn{2}{c}{$4\times1$\,pix}\\
Virtual pixel scale & $9\farcs36\times9\farcs36$ & $10\farcs4\times9\farcs36$ \\
Detection limit (5$\sigma$) & 50\,mJy$^{\dagger}$ & 120\,mJy$^{\dagger}$\\
\hline
\end{tabular}\\
$*$ Effective exposures for point sources are determined not by the sampling rate
but by the dwelling time of a source on a pixel.\\
$\dagger$ Estimated value from readout noise in shuttered configuration in-orbit.\\
$\ddagger$ Defined as where the responsivity for a given energy is larger than
$1/e$ of the peak.
\end{table}

\paragraph{Array Operation}
During the all-sky survey observations, 
only two out of 256 rows in the sensor array
are operated
in the continuous and non-destructive readout mode
\citep[scan operation of the array;][]{ScanOpe}.
The 1$^{\rm st}$ row used in the operation
can be selected arbitrarily out of 256 rows
and we adopted 117th row (hereafter row\#1).
The 2$^{\rm nd}$ row (hereafter row\#2)
was fixed to be 8 rows from the 1$^{\rm st}$ row
the sampling rate of the array and
the designed scan rate of the satellite.

The sampling rate was set to 22.72\,Hz (one sampling per 44\,ms)
taking account of the array operation conditions
and the data down-link capacity.
All the pixels are reset at a rate of 0.074\,Hz (one reset per 306 samplings)
to discharge the photo-current.

The NIR channel is not used during the all-sky survey 
because of the capacity of the down link rate
and because
the alignment of the NIR array is not suited for the all-sky survey
observation.

\paragraph{Exposure time}
The scan speed of the satellite in the survey observation mode is 216$'' {\rm s}^{-1}$.
As the pixel scale of the detector in the in-scan direction is about $2\farcs34$, 
the resulting effective exposure time for a point source is 11\,ms.

\begin{figure}
\includegraphics[width=8cm]{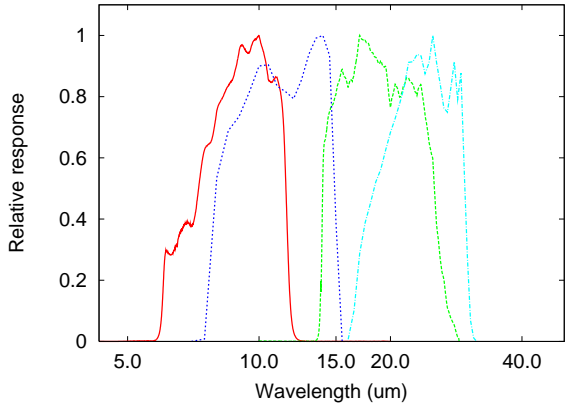}
\caption{Relative spectral response curve of the S9W (solid curve)
and L18W bands (dashed) in units of electron/energy
normalized to the peak.
The system response curves of the IRAS 12\,$\mu$m (dotted)
and 25\,$\mu$m bands (dashed-dotted)
are also shown for comparison.
The RSRs of the IRC are available at
http://www.ir.isas.jaxa.jp/ASTRO-F/Observation/RSRF/IRC\_FAD/index.html. 
The system response curves of the IRAS are taken from
Table II.C.5 of IRAS Explanatory Supplement \citep{IRASsupp}.
}
\label{fig:rsr}
\end{figure}

\paragraph{Filter bands}
The all-sky survey is performed with two broad bands
centered at 9\,$\mu$m (S9W filter of MIR-S) and 18\,$\mu$m (L18W filter of MIR-L).
The relative spectral response (RSR) curves of the two bands are shown
in Fig.~\ref{fig:rsr} together with those of the {\it IRAS} 12 and
25\,$\mu$m bands.

\paragraph{Pixel scale}
The effective pixel size in the survey observation (hereafter virtual pixel size)
is $9\farcs36\times9\farcs36$ for 
the 9\,$\mu$m band, and $10\farcs4\times9\farcs36$ for the 18\,$\mu$m band.
The in-scan pixel size is fixed by the sampling rate,
while the outputs of 4 neighboring pixels are coadded 
to meet the down-link rate requirements and still make effective observations.
A finer resolution is obtained in the data processing
by combining images produced by the two rows.
A detailed description of the reconstruction of
the image and the resulting spatial resolution are given in Appendix~\ref{nest4x4}.

\paragraph{Confirmation strategy}
It is difficult to distinguish
events of real celestial objects 
from those due to cosmic ray hits 
only from the shape of signal
because the virtual pixel size is not small enough 
to sample the PSF.
We improve the reliability of source detections
by adopting a three-step confirmation scheme.

The detection of a celestial source in row\#1 is confirmed
by a 2$^{\rm nd}$ detection in row\#2 87\,milli-seconds later.
In this process, a large fraction of the signals due to
cosmic ray hits are
expected to be removed (milli-seconds confirmation).

The width of the array in the cross-scan direction is about 10$'$.
The scan path shifts at most by 4$'$ (on the ecliptic plane)
due to the orbital motion.
Thus, the scan path overlaps at least by 6$'$ 
with the next scan 100 minutes after.
The detection is thus comfirmed by the next scan osbervation.
(hours confirmation).

Furthermore, the scan path rotates by 180 degrees
around the axis of the earth in half a year,
giving another chance of detection six months later (months confirmation).
Objects, such as asteroids, comets and geostationary satellites,
can be distinguished from stars and galaxies 
by hours- and/or months- confirmations.

\paragraph{Sky coverage}
Large portions of the sky have a chance to be covered more than three times
during the survey period (Phase1, Phase2a, and Phase2b; \S\ref{akari}).
During the semi-continuous survey,
observational gaps (where no data are effectively available) appear
because:
(1) The survey observations are halted for every pointed observation.
A sky area of about $10'\times120^\circ$ is skipped during a pointed observation.
(2) During Phase 2 the attitude of the satellite was operated actively also in the
all-sky survey mode to cover the sky area not observed in 
Phase 1 in order to complete the far-infrared FIS all-sky survey (offset survey).
However it is the intermission of the continuous survey 
for the mid-infrared channels which have FOV directions different from the FIS.
(3) The mid-infrared survey observations are 
sometimes halted because of the limited downlink capacity.
(4) The attitude of the satellite is lost due to a star tracking failure.
(5) The shutter is closed in the moon avoidance region.
(6) The data taken during resets and during times affected
by the heavy reset anomaly are masked by the pipeline software (\S\ref{data:basic}).
(7) Saturated pixels can not observe the sky until
having been reset (saturation trail).

\begin{figure}
\center
\includegraphics[width=8.8cm]{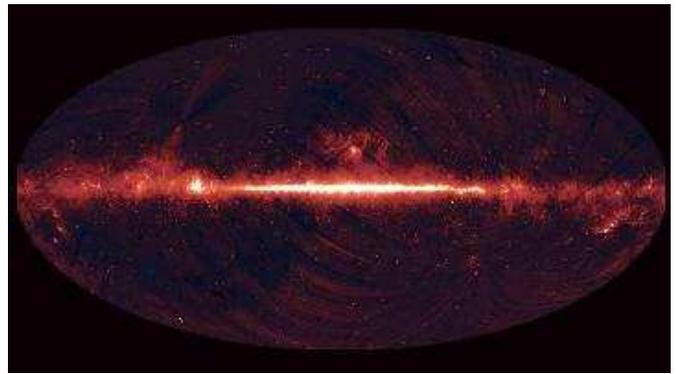}
\caption{All-Sky image taken with the {\it AKARI} 9\,$\mu$m band,
from which the zodiacal light is subtracted
simply as the low spatial frequency component
derived from the raw data.
}
\label{fig:diffusesky}
\end{figure}

Taking into account (1), (2), and (3),
which are automatically detected from the telemetry data,
the final sky coverage 
is higher than 90\% for both bands.
Note that the actual sky area in which the source
confirmation is carried out is smaller than this number (see \S\ref{cat_compile}).
Fig.~\ref{fig:diffusesky} shows the
{\it AKARI} 9\,$\mu$m low-resolution intensity map
derived through the data processing described in this paper,
from which the zodiacal light has been subtracted.

\section{Data processing}

\begin{figure*}
\center
\includegraphics[width=9cm]{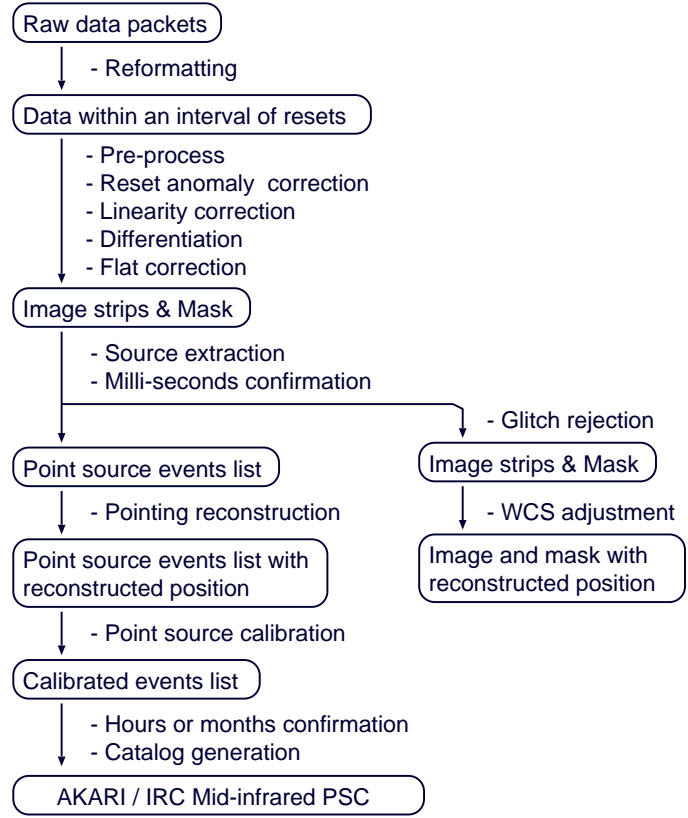}
\caption{Outline of {\it AKARI}/IRC all-sky survey data processing.}
\label{fig:outline}
\end{figure*}

The outline of the data processing (Fig.~\ref{fig:outline})
for deriving the first point source catalogue
of the {\it AKARI}/IRC mid-infrared all-sky survey is summarized in the following:
\begin{itemize}
\item The raw telemetry data, which include
the output of the sensor array, 
the house keeping data, and the output of the attitude control system,
are down-linked from the satellite. Their timing is
matched with each other and then they are registered to the database.
\item The raw data that are sandwiched by
two successive resets in the scan are
processed
to make pieces of images corresponding to $10'\times50'$ wide sky
regions (hereafter unit images) (Basic process; \S\ref{data:basic}).
\item Then signals of the point source detections (events) are extracted
from each processed image
(Event detection and milli-seconds confirmation; \S\ref{data:ext}).
\item Next, the coordinates of all the events are determined
according to the output of the attitude control system
and refined using cross-correlations between the
detected events and the prepared standard stars for
the position determination
(Pointing reconstruction; \S\ref{PR}).
\item After the absolute flux calibration based on the measurements of standard stars,
the fluxes of all the detected events are statistically derived (Flux calibration; \S\ref{data:flux}).
\item Finally, reliable events are compiled into point source lists and the source lists of the 9\,$\mu$m
band and 18\,$\mu$m band are merged together
to produce the IRC all-sky survey catalog (Catalogue compilation; \S\ref{cat_compile}).
\end{itemize}

\subsection{Basic process \label{data:basic}}

The basic process derives
a unit image of $10'\times 50'$ from the raw data
for the extraction of point source candidates.

\paragraph{ADU to electron conversion}

First,
the pixel value of the raw data is converted from ADU into electrons as
\begin{equation}
 S^{(1)}(i,t) = CF \cdot S^{(0)}(i,t) ,
\end{equation}
where $i$ is the pixel number, $t$ is the time from the latest reset,
$S^{(n)}(i,t)$ is the pixel value at $(i,t)$ as the result of applying the
$n$th step, and $CF$ is the conversion factor 
measured from laboratory tests 
and assumed to be a constant for the pixels in the 
detector array \citep{Ishihara03}.

\paragraph{Reset anomaly correction}

\begin{figure}
\includegraphics[width=8cm]{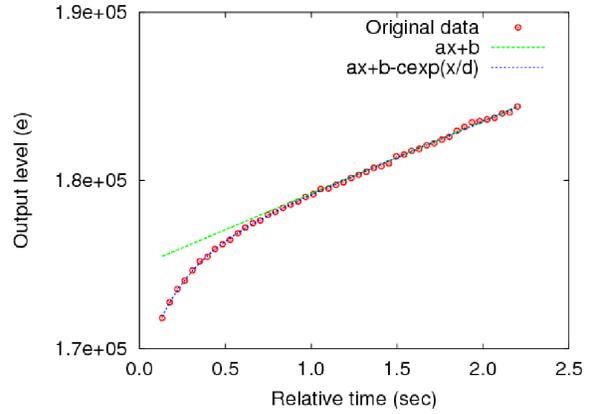}
\caption{Example of the reset anomaly correction.
The output signal level of a pixel in units of electron
is plotted against the time from latest reset.
This data set was taken under the constant illuminating source
in the laboratory test.
The raw output of the sensor (solid points)
is fitted by the reset anomaly function $O(t,e_{\rm int})$ (dotted curve)
and corrected (dashed line).
}
\label{fig:rano}
\end{figure}

An anomalous behavior of the output level
that persists a few seconds after the reset (reset anomaly) is corrected.
An example of this phenomenon and its correction are shown in Fig.~\ref{fig:rano}.
The offset level of the output of the detector is fairly sensitive to the
temperature \citep{Ishihara03}.
We thus suppose that this phenomenon
is explained by the drift of the offset level of the read-out circuit
hybridized to the detector array
owing to the temperature drift
invoked by the reset current to discharge the stacked photo-electrons.
Assuming that this behavior is pixel-independent in the detector array,
the reset anomaly is corrected as
\begin{equation}
 S^{(2)}(i,t) = S^{(1)}(i,t) - O(t, e_{\rm{res}}) ,
\end{equation}
where $i$ is the pixel number, $t$ is the time from latest reset, 
$e_{res}$ is (pixel average of) the amount of the
photo-electrons discharged during the latest reset,
and $O(t,e_{res})$ is the offset level variation.
The offset level variation $O(t,e_{res})$ due to the reset
is represented in the form of
\begin{equation}
 O(t, e_{int}) = C_1 \exp{(-t / C_2)} ,
\end{equation}
where $C_1$ and $C_2$ are constants statistically 
derived from laboratory tests.

\paragraph{Masking invalid data}

A masking pattern is created to ignore unusable data.
(1) All the pixels are masked during the
resets and the periods heavily affected by the reset anomaly.
(2) Saturated pixels are masked 
from the moment of the saturation to the next reset.
(3) The pixels under the slit masks are always masked.
Both MIR-S and MIR-L channels have a slit and 
slit masks at the edge of the FOV for spectroscopic observations
of extended objects.
The slit mask covers a few pixels located near the edge
of the rows used in the survey operation.

\begin{figure}
\includegraphics[width=7.5cm]{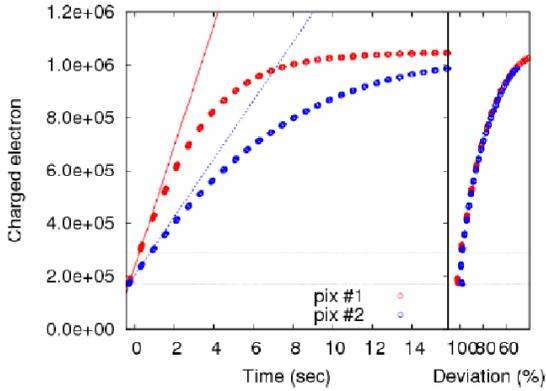}
\caption{Example of the non-linearity of the photo-response of a pixel.
Left panel shows time from reset versus output level of two pixels
at the constantly illuminated condition in a laboratory test.
The output profiles are fitted by linear functions
using the data in the range, where the photo-response is still linear
because the number of stacked electrons is small (7000--15000 ADU).
Right panel shows the number of charged electrons (Y-axis)
versus the deviation of the output level from the linear functions (X-axis).
}
\label{fig:lin-sample}
\end{figure}

\paragraph{Linearity correction}
We correct for the non-linearity of the photo-response ($L(e_{\rm int})$),
which is supposedly
due to the decrease in the bias voltage imposed on the 
detector array
by the accumulation of the photo-electrons in each pixel.
It is supposed to be a function of
the number of stacked electrons ($e_{\rm int}$).
An example of the non-linearity is shown in Fig.~\ref{fig:lin-sample}.
The non-linearity of the photo-response is thus corrected as
\begin{equation}
S^{(3)}(i,t) = S^{(2)}(i,t) / L(e_{\rm int}),
\end{equation}
assuming that their behaviors are the same for all the pixels.
The non-linearity function $L(e_{\rm int})$ is approximated
by a spline function:
\begin{equation}
L_j(e_{\rm int}) = \sum_{i=0}^{3} C_j^{(i)} e_{\rm int}^i ,
\end{equation}
where $j$ ($j$=0,1,2) represents the $j^{\rm th}$ range of $e_{\rm int}$
and each coefficient ($C_j^{(i)}$) is derived from laboratory tests.

\paragraph{Differentiation}
The signal is differentiated with respect to time,
\begin{equation}
 S^{(4)}(i,t) = S^{(3)}(i,t+\Delta t) - S^{(3)}(i,t) ,
\end{equation}
where $\Delta$t is the time step (44\,ms) corresponding to the sampling rate.

\begin{figure*}
\center
\includegraphics[width=7cm]{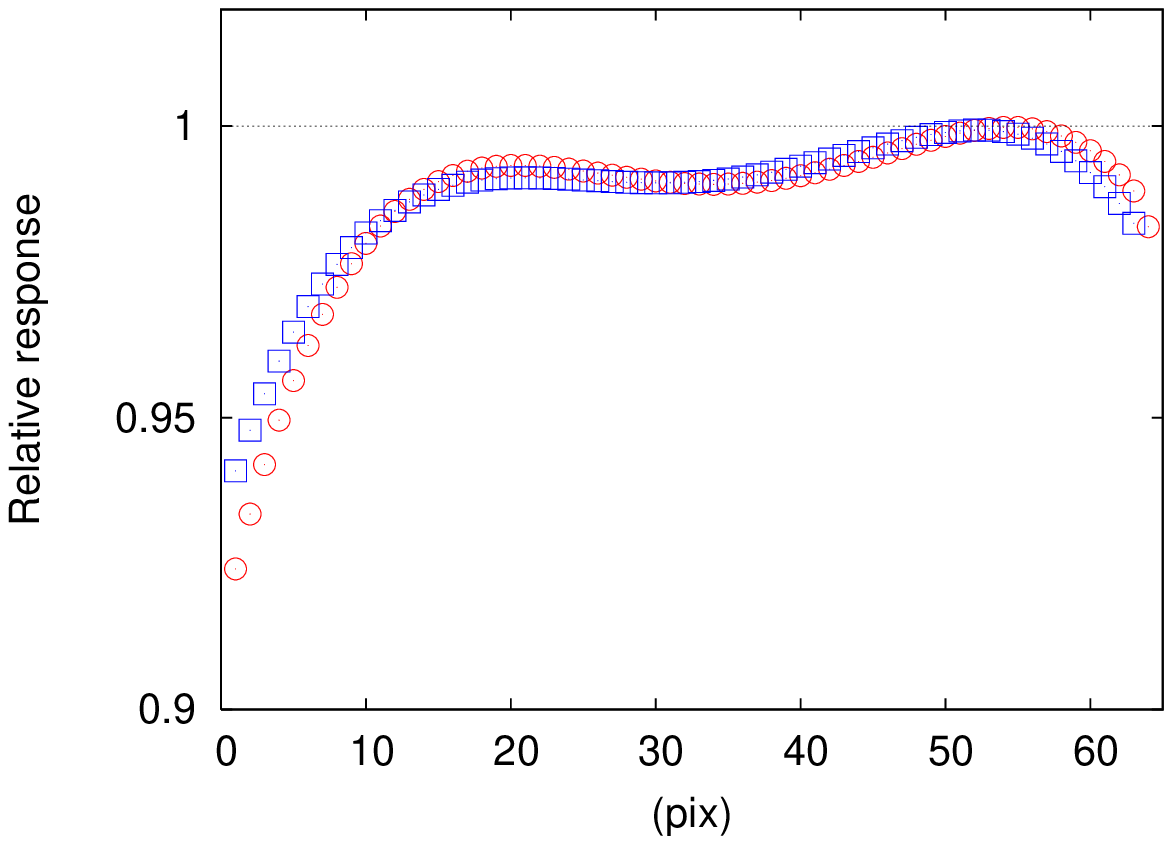}
\includegraphics[width=7cm]{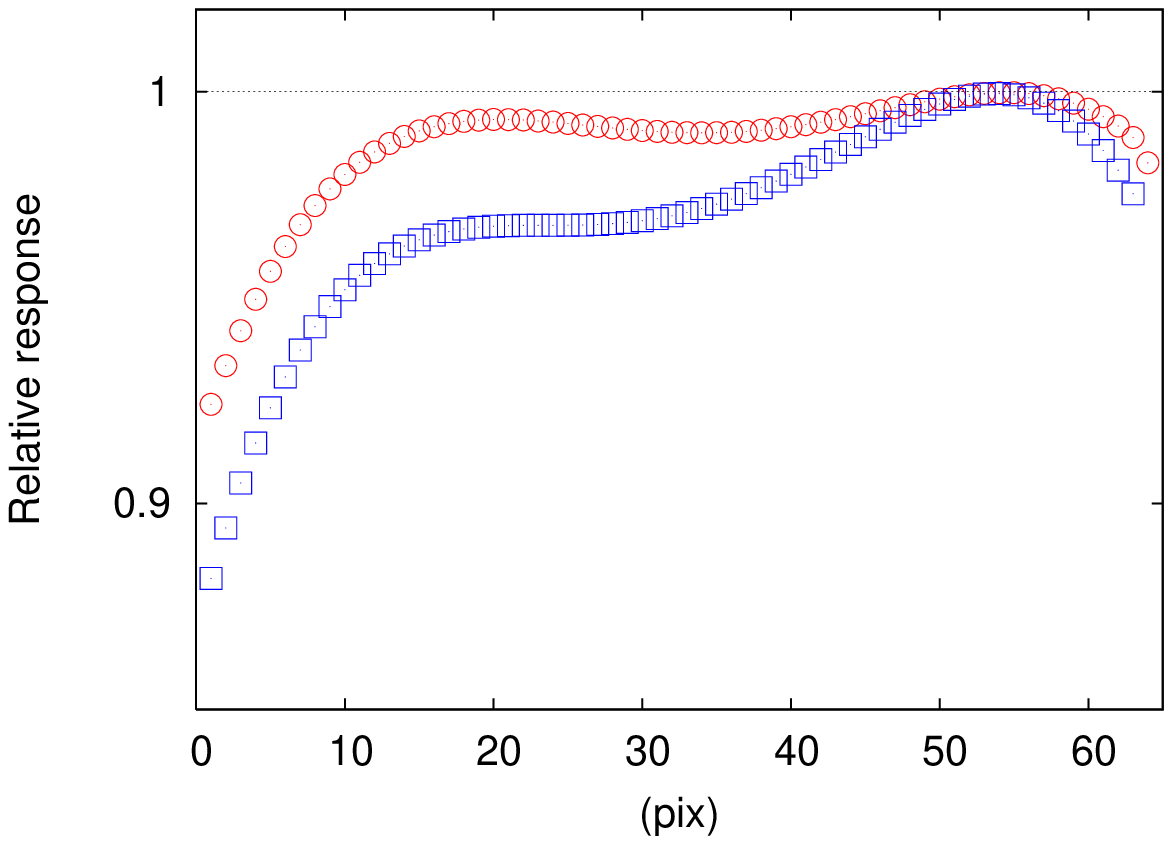}
\caption{Flat functions for the one dimensional array
for the 9\,$\mu$m band (left) and 18\,$\mu$m band (right).
The open circles represent row\#1 and 
open squares represent row\#2 in both panels.
}
\label{fig:flat}
\end{figure*}

\paragraph{Flat fielding}
The differentiated signal $S^{(4)}(i,t)$ is corrected for flat-fielding.
\begin{equation}
 S^{(5)}(i,t) = S^{(4)}(i,t) / F(i) ,
\end{equation}
where $F(i)$ is the normalized flat correction factor for each pixel
operated in the survey mode
representing the dispersion of the photo-response among the pixels.
The flat function $F(i)$ is derived from multiple detections of the stars 
on different pixels.  The background sky data are not used
to avoid the effects of the scattered light.
Fig.~\ref{fig:flat} shows the flat functions for both rows in both bands.

\paragraph{Image construction from data of the two rows}

\begin{figure}
\includegraphics[width=8cm]{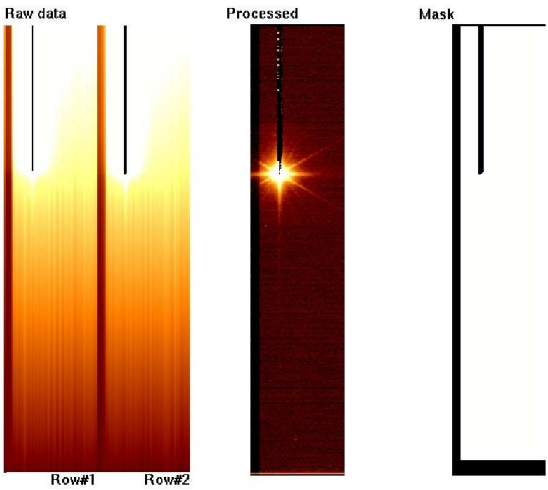}
\caption{Example of the processed data. 
(left) Raw data from Row\#1 and Row\#2 are shown separately.
(middle) Processed data. The data from both rows are combined.
(right) Mask for the processed data.}
\label{fig:sample}
\end{figure}

The data from row\#1 and row\#2 are combined
into a single image with a pixel scale of $1\farcs56$
to reject cosmic ray hits and
produce a finer and higher S/N image.
Details on the combining process are given in Appendix \ref{nest4x4}.
The mask patterns from the two rows are also combined in the same manner.
Fig.~\ref{fig:sample} shows examples
of the raw data obtained by the two rows,
the processed data of the two rows,
(i.e. the resulting combined image of $10'\times50'$), and
the combined mask pattern.
The images and mask patterns are
stacked as the basic calibrated data
and put into the next step, the source extraction process.

\subsection{Event detection and milli-seconds confirmation \label{data:ext}}

\begin{figure*}[h]
\center
\includegraphics[width=12cm]{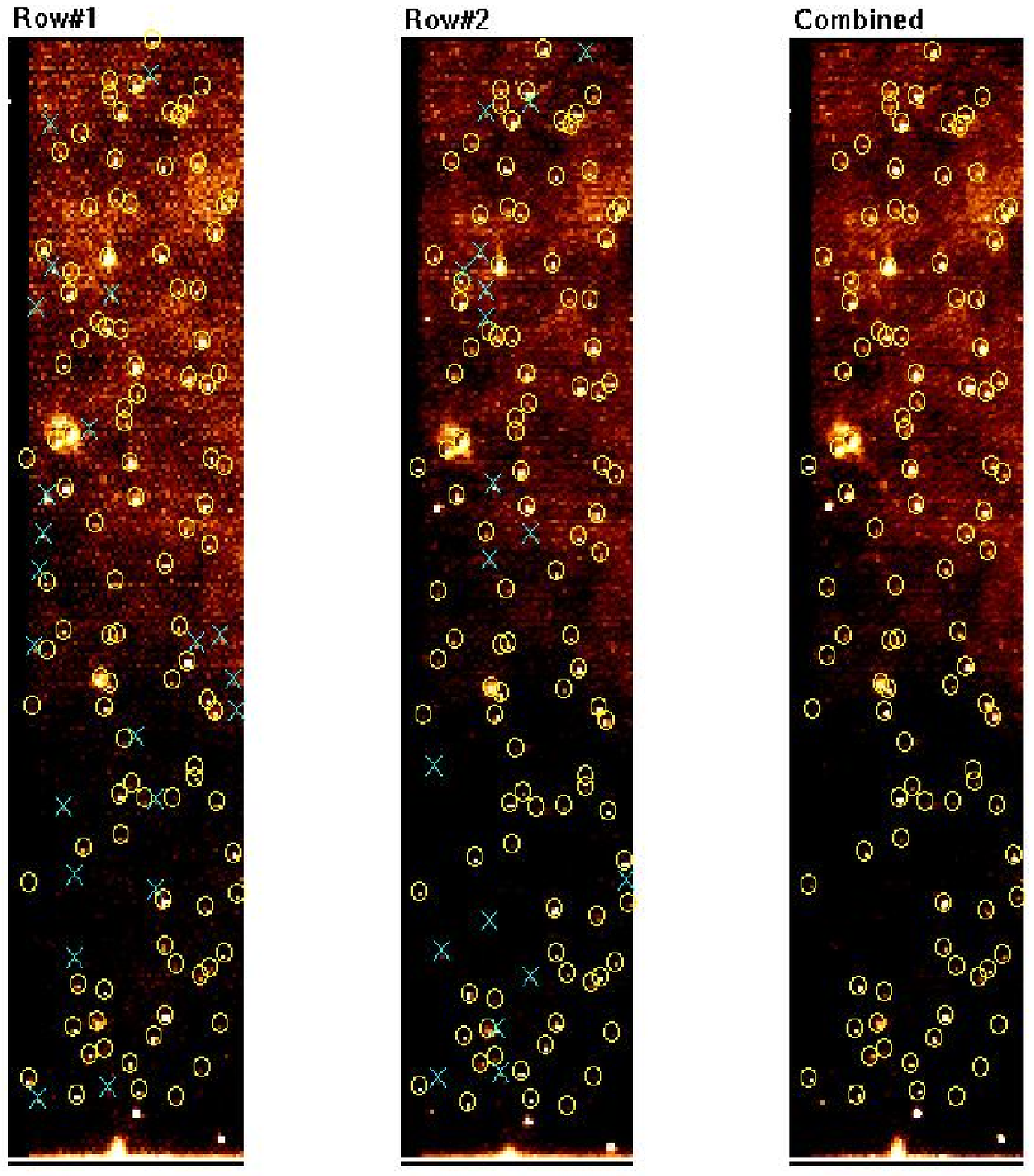}
\caption{Example of the event detection and milli-seconds confirmation.
The left image is obtained by row\#1 and the center image is obtained by row\#2.
The right image is constructed by combining data from two rows.
The crosses mark events
extracted on a single row image but rejected in milli-seconds confirmation.
The circles show events extracted on both rows with similar fluxes and thus confirmed.}
\label{fig:scon}
\end{figure*}

After the basic processing, 
signals 
above 5\,$\sigma$ per scan
are extracted from each image and checked by milli-seconds confirmation
for point source detection (hereafter event).
An example of these processes is shown in Fig.~\ref{fig:scon}.

First, events are extracted from two images
obtained independently by the two rows,
using the Source Extractor \citep{SExtracter} with a 3$\sigma$ threshold.
Then events above 3$\sigma$ are extracted again on the finner combined image.
Events from row\#1, row\#2 and the combined image are 
cross-identified by the detected position.
Celestial sources are expected to be detected in both rows
on the same pixel number ($i$) with similar fluxes,
whereas false detections such as cosmic ray hits are 
expected to be detected only on one side.
A pair of detections at the same sky position
with similar fluxes ($0.1 < F_{\rm row\#1} / F_{\rm row\#2} < 10$) are
selected as milli-seconds confirmed sources.
Finally, the flux (photometric result)
and the position (timing and pixel number of the detection)
of the combined image are recorded for the confirmed events.

\subsection{Pointing reconstruction \label{PR}}

The celestial coordinate of a source ($Q_{\rm pix}(i,t)$) 
detected on the $i$-th pixel at the moment $t$ are
derived from
\begin{equation}
Q_{\rm obj}(i,t) = L(t) E(i) Q_B(t) ,
\end{equation}
where $Q_B(t)$ is the boresight position of the telescope
at the moment $t$,
$E(i)$ is the conversion from the boresight to the $i$th pixel,
and 
$L(t)$ is the correction for the aberration due to
the yearly revolution of the Earth and the orbital motion of the satellite.

The improvement of boresignt position of the telescope $(Q_B(t))$
is carried out
in collaboration with ESA
by associating 
detected events in the 9 and 18\,$\mu$m bands
and signals of the focal-star sensors at near-infrared on the focal plane
with the stars in the positional reference catalogue prepared from {\it MSX}, {\it 2MASS}
and {\it IRAS} (pointing reconstruction).
The actual alignment of the FOV of each pixel $E(i)$ including
the effect of the distortion
is optimized statistically in the pointing reconstruction process.

\begin{figure}
\includegraphics[width=8.5cm]{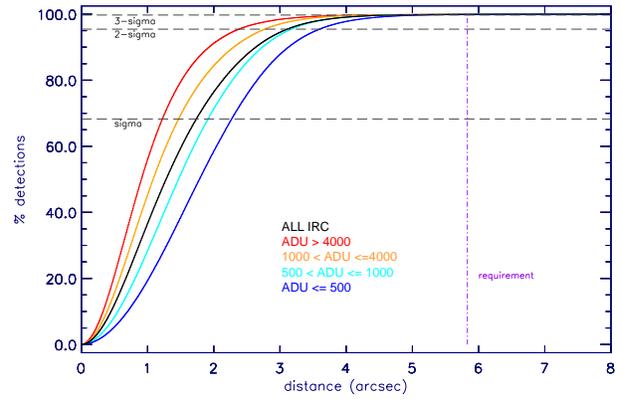}
\caption{Statistical error of the pointing reconstruction using 50\% of the
catalogue data. The error 
is defined as
the distance between the position determined by the pointing reconstruction and
the position of the input reference catalogue for events in the remaining 50\% data.
The lines show the fraction of the IRC events with an error smaller than
given values. The color of the lines denotes the flux range of events.}
\label{fig:fig-pr}
\end{figure}

Fig.~\ref{fig:fig-pr} shows the accuracy of the
pointing reconstruction.
The plot shows the fraction of the IRC events with an error
smaller than given values.
The error is estimated from the distance between the positions determined from
the pointing reconstruction and those from the position
reference catalogue.
To make fair evaluation, the pointing reconstruction for this test is carried
out using randomly selected sources amounting to half of the catalogue, and
then the positions of the sources from the other half of the
catalogue are determined
for the evaluation.
The error includes the pointing reconstruction processing and
the measurement errors. For the brightest sources the measurement error
should be a minor contributor.  We
conclude that the position accuracy is better than
$3\arcsec$ for 95 \% of the events.

Details of the pointing reconstruction are summarized in Salama et
al. (2010). 
We adopt the position of each detected event
derived from the pointing reconstruction process.

\subsection{Flux calibration \label{data:flux}}

The photometric output for all the milli-seconds confirmed events
is converted from electrons into physical units (Jy).
The conversion function from electrons to Jy is derived statistically
by comparing the model fluxes with
the measurements of hundreds of standard stars.
The standard stars are selected
from the infrared standard star network consisting of K- and M-giants
\citep{Cohen09}
and additional faint standard stars
located around the north and south ecliptic poles \citep{Reach05,CohenFaint},
which have a high visibility for the {\it AKARI} survey.
The expected fluxes ($f^{\rm quoted}_\lambda(\lambda_i)$)
of the standard stars at the effective wavelengths (9 and 18\,$\mu$m) 
are calculated for the incident spectrum of $f_\lambda \propto \lambda^{-1}$
by convolving the model spectra of the standard stars ($f_\lambda(\lambda)$)
with the relative spectral response curves ($R_i$) in electron units 
as
\begin{equation}
  f_\lambda^{\rm quoted} (\lambda_i) =
	 \frac
	{\int R_i(\lambda) \lambda f_\lambda(\lambda) d\lambda}
	{\int (\frac{\lambda_i}{\lambda})R_i(\lambda) \lambda
	d\lambda} ,
\end{equation}
where $i$ represents the band $i$, $\lambda_i$ is the effective wavelength of the
band.

The fitting function to convert
the measured signals $P$ into fluxes $F$ is given by
\begin{equation}
 \log(F) = \sum_{i=0}^2 C_i \log(P)^i ,
\end{equation}
where $C_i$'s are fitting coefficients
(their values are given in the release note).
Finally, the derived conversion functions are applied to
all the milli-seconds confirmed events.
The zero magnitude flux is
$54.5\pm0.845$\,Jy and $12.3\pm0.204$\,Jy 
for the 9\,$\mu$m and 18\,$\mu$m bands, respectively.

\begin{figure*}
\center
\includegraphics[width=8cm]{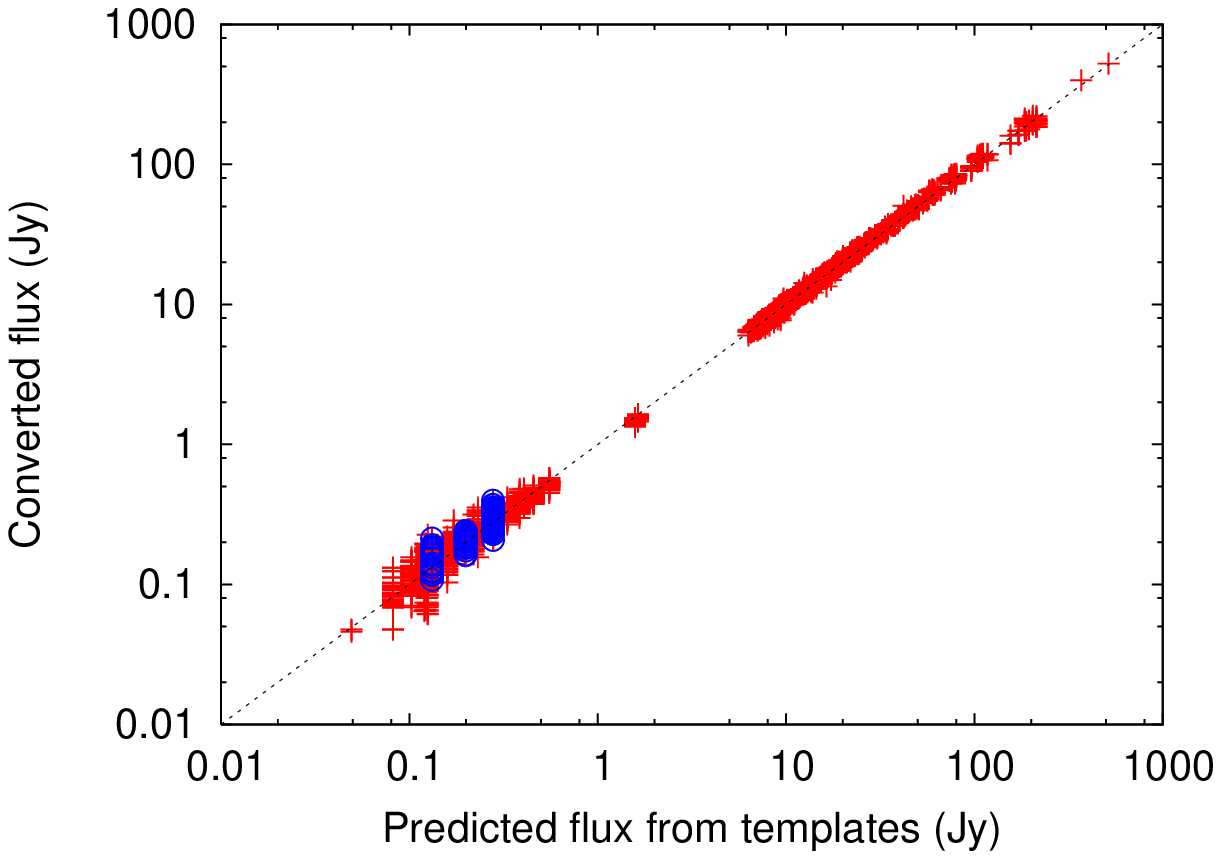}
\includegraphics[width=8cm]{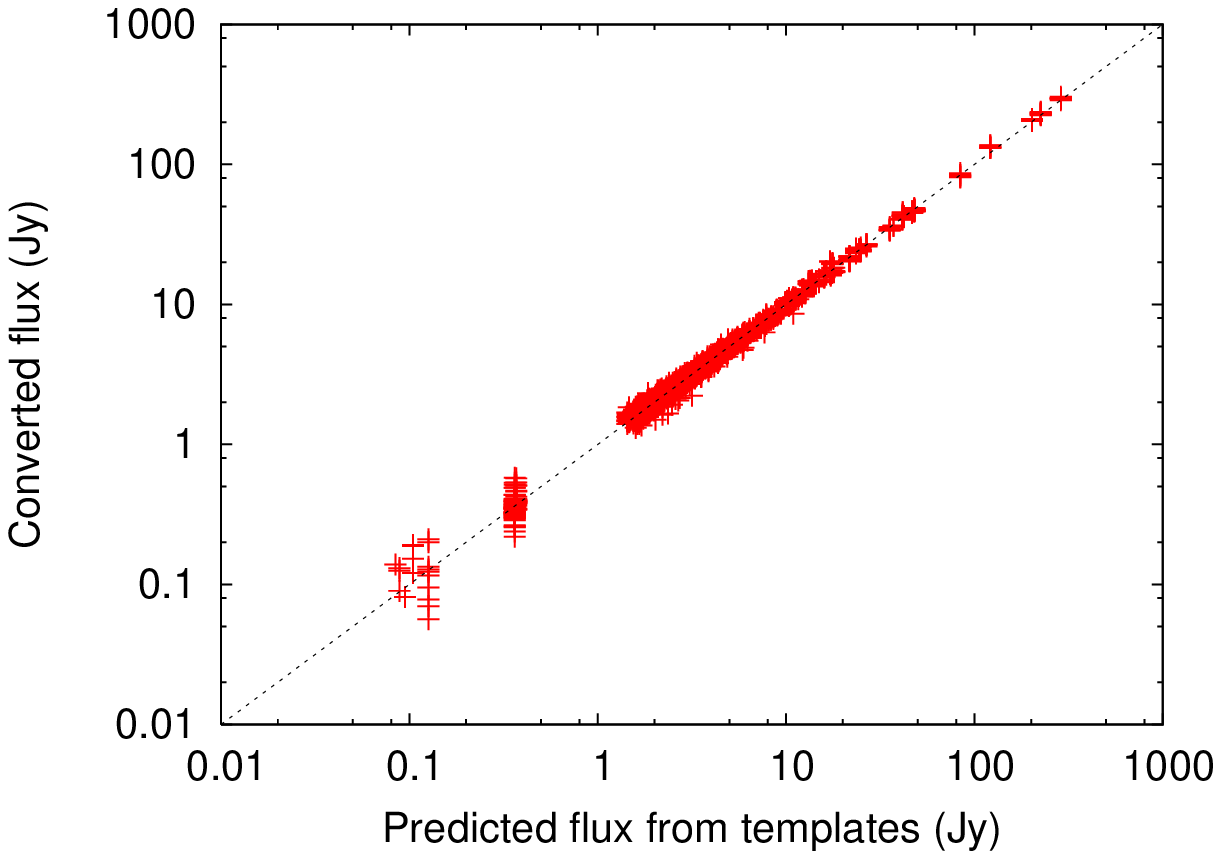}\\
\includegraphics[width=8cm]{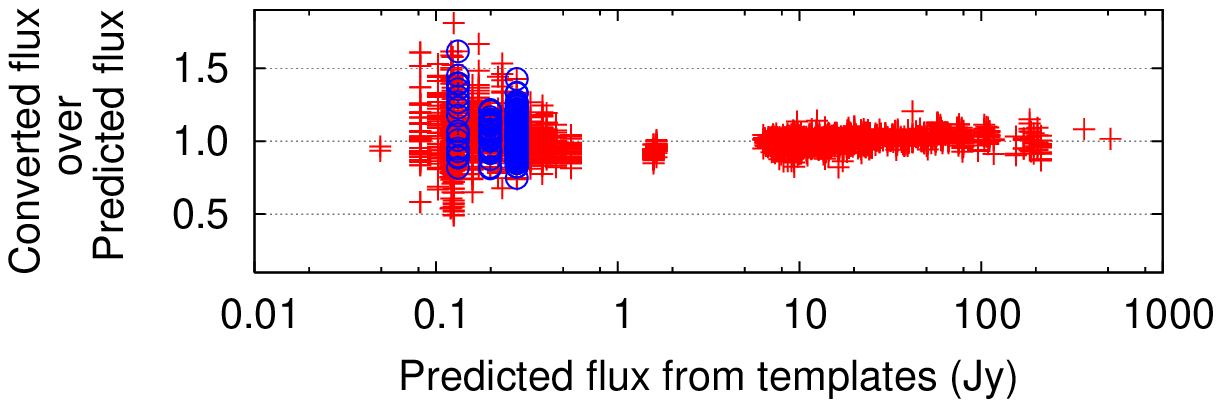}
\includegraphics[width=8cm]{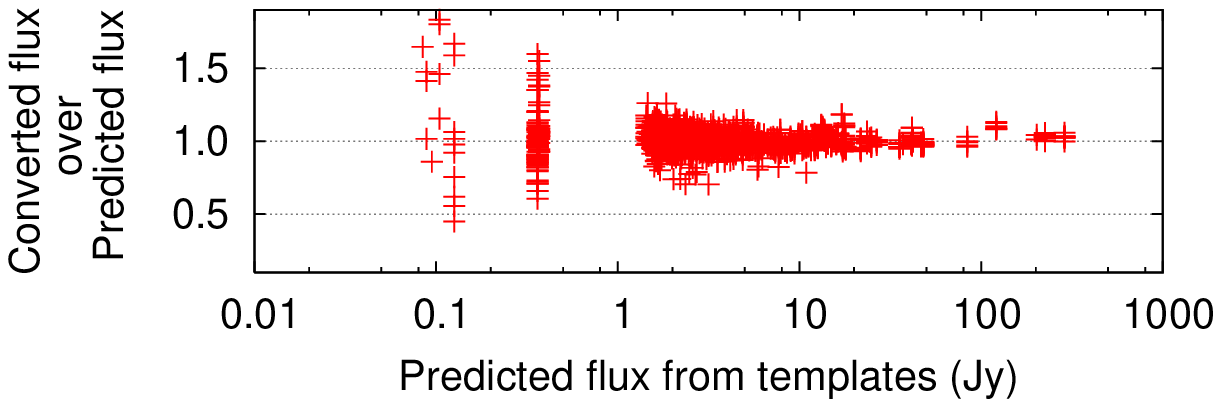}\\
\caption{
(Top left) Calibrated fluxes obtained from the
pipeline output through Eq. (10) are plotted as a function of the
predicted in-band fluxes  of the standard stars
in the 9\,$\mu$m band.  The slid line indicates $y=x$ for
reference.  (Bottom left) Ratio of the measured to
predicted fluxes for the the standard stars.  The right
panels show the same plots for the 18\,$\mu$m band.  The symbols and
lines are the same as in the left panels.  
The blue points show three A-type stars
used for the 9\,$\mu$m band calibration.
All the standard stars used for the 18\,$\mu$m band calibration
are K--M giants and no A-type stars are used.
}
\label{fig:fcal}
\end{figure*}

Fig.~\ref{fig:fcal} (top) shows the flux derived
(by Eq. 10) from the pipeline output
as a function of the predicted in-band flux of the standard stars.

The accuracy of the calibrated flux is investigated by
the ratio of  the measured flux to the predicted flux of the
standard stars. The results shown in the bottom panels of Fig.~\ref{fig:fcal} 
indicate an accuracy for the absolute calibration
of about 3\% for the 9\,$\mu$m band and
4\% for the 18\,$\mu$m band.

A systematic offset between A-type and K-M type giant standard
stars is reported by Reach et al. (2005)
for the calibration of the IRAC on {\it Spitzer}.  We have only three
A-type standard stars in our calibration for the 9\,$\mu$m band and
no A-type stars are used for the calibration of the 18\,$\mu$m band
due to the detection limit.  No systematic offset is seen in the
present calibration between A-type stars and K-M giants within the
measurement uncertainties (Fig.~\ref{fig:fcal}).  Note that the calibration
of pointing observations of the IRC imaging mode does not show
any appreciable offsets either \citep{Tanabe}.

\begin{figure}
\includegraphics[width=8.5cm]{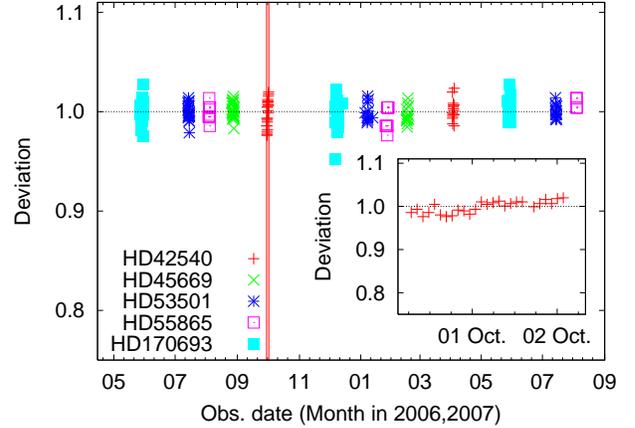}
\caption{
Ratio of the measured fluxes to the average fluxes
as a function of observing time for five bright ($>1$\,Jy) standard stars
observed at 9\,$\mu$m more than 30 times during the survey.
The stars used and the corresponding 9\,$\mu$m fluxes and
number of detections are:
HD42540 (plus, 8.33\,Jy, 44\,times), HD45669 (cross, 9.79\,Jy, 43\,times),
HD53501 (star, 9.30\,Jy, 45\,times), HD55865 (open box, 16.7\,Jy, 33\,times)
and HD170693 (filled box, 9.60\,Jy, 76\,times), respectively.
The closeup shown in the right bottom panel is to check for the
temporal variation on a short timescale.
}
\label{fig:stable}
\end{figure}

To test of the long-term stability of the photo
response we investigate five standard stars 
that have been observed more than 30 times
during the time of the survey. Fig.~\ref{fig:stable} shows the ratio of the
fluxes of individual measurements
to the average fluxes of these stars as a funcion of
time. From these data we deduce that the sensitivity is stable at
the $\sim$2\% level during the entire period of the observations.

Laboratory measurements of the filter transmission indicate possible blue
leaks between 3 and 4\,$\mu$m for the 9\,$\mu$m band
and between 6 and 7\,$\mu$m for the 18\,$\mu$m band.  
They are 0.01\% at maximum
and much smaller than the measurement errors.
Thus the presence of the blue leaks is not confirmed.
We calculate the predicted fluxes of the standard stars
with and without the blue leak and confirmed that 
the difference is less than 0.1\%.  
We have verified
that the blue leak of 
the 18\,$\mu$m band
is negligible ($<1$\%)
compared to the
systematic errors by using asteroid calibrators whose flux can be well
predicted (M\"{u}ller \& Hasegawa, private communication).

\subsection{Catalogue compilation \label{cat_compile}}

The final list of the sources  
(hereafter Point Source Catalogue; PSC)
has been prepared based on the following criteria.
First, groups of multiple events located within a region of $5\arcsec$
radius are recognized as a same source. In this process, events
in the South Atlantic Anomaly (SAA) are excluded.  Only
groups containing at least two events are recognized as an actual celestial
source.
After the first grouping process, 
the distance from a source candidate to the nearest one
is investigated and if there are two or more groups within $7\arcsec$,
we consider them as a single source associated with outskirts events.
In this case we take the group with the maximum
number of events as a source candidate and discard the other groups.
Then, the source lists in the 9\,$\mu$m 
and 18\,$\mu$m band are merged into a single list. 
Sources within 7$''$  are regarded as the same source in both bands.
The position (RA, DEC) and associated position error
(the major and minor axes and the position angle) are calculated from the events
in the 9\,$\mu$m band
only, if the number of available events is larger than or equal to 2.
Otherwise, these data are calculated from the events in the 18\,$\mu$m band only.
The flux and associated error
of the source are estimated from
the mean and the mean error of 
multiple measurements of the events, respectively.
Event data near the edge of the image strips are excluded
from the flux calculation
unless the exclusion leaves
only zero or one event.

\begin{figure}
\includegraphics[width=9cm]{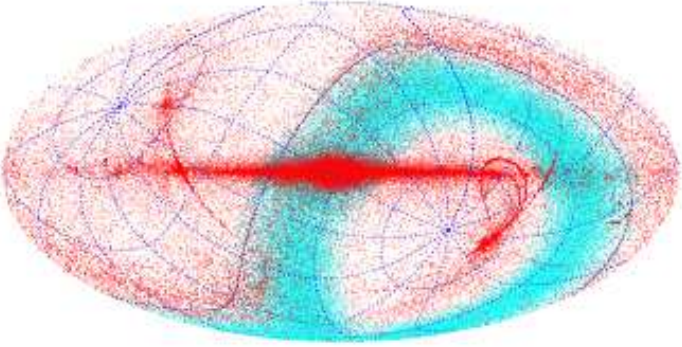}
\caption{ Spatial distribution in Aitoff Galactic coordinates
projection of the 18\,$\mu$m events rejected in the catalogue compilation
process.  The grids indicate the equatorial coordinates.  Most of
the events correspond to  high energy particles due to the
South Atlantic Anomaly (SAA), or to  moving objects such as asteroids and
comets (aligned around $|\beta|<10^\circ$), to geostationary satellites
($|{\rm Dec}|\sim 0^\circ$), and to retired geostationary satellites
($|{\rm Dec}|<10^\circ$). The colour version will be available in the
online version, in which the SAA events are plotted in cyan, and the
remaining objects in red.
}
\label{fig:rej}
\end{figure}

The numbers of the events obtained in each step of the process
are summarized in Table \ref{tbl:event}.
The spatial distribution of the rejected events 
(those without hours and months confirmation) is shown in Fig. \ref{fig:rej}.
Most of the rejected events are ascribed to asteroids,
geostationary satellites, and the high-energy
particle hits in the SAA.
These objects are recognized as such
thanks to 
their spatial distribution and characteristics features.
It should be noted that the actual sky coverage
for this catalogue is smaller than the value
quoted in \S\ref{sec:scanope};
this is due to
the severe condition of $N_{\mathrm{events}} \ge 2$
as well as to the exclusion of all data affected by the SAA.

\begin{table}
\caption{Number of events at each step of the process for the preparation of the PSC. \label{tbl:event}}
\begin{tabular}{lrr}\hline\hline
  & 9\,$\mu$m & 18\,$\mu$m\\\hline
Detected                  & 16,752,471 & 9,655,059 \\
Milli-seconds confirmed   & 4,929,586 & 1,300,945\\
Hours or Months confirmed & 851,189$^*$ & 195,893$^*$\\
Band merged               &\multicolumn{2}{c}{877,091$^*$}\\ \hline
\end{tabular}\\
$*$ -- Current values. The number of sources 
may be updated in further catalogue releases.
\end{table}

\section{Evaluation of the catalogue}

\subsection{Spatial distribution}
\begin{figure*}[h]
\includegraphics[width=19cm]{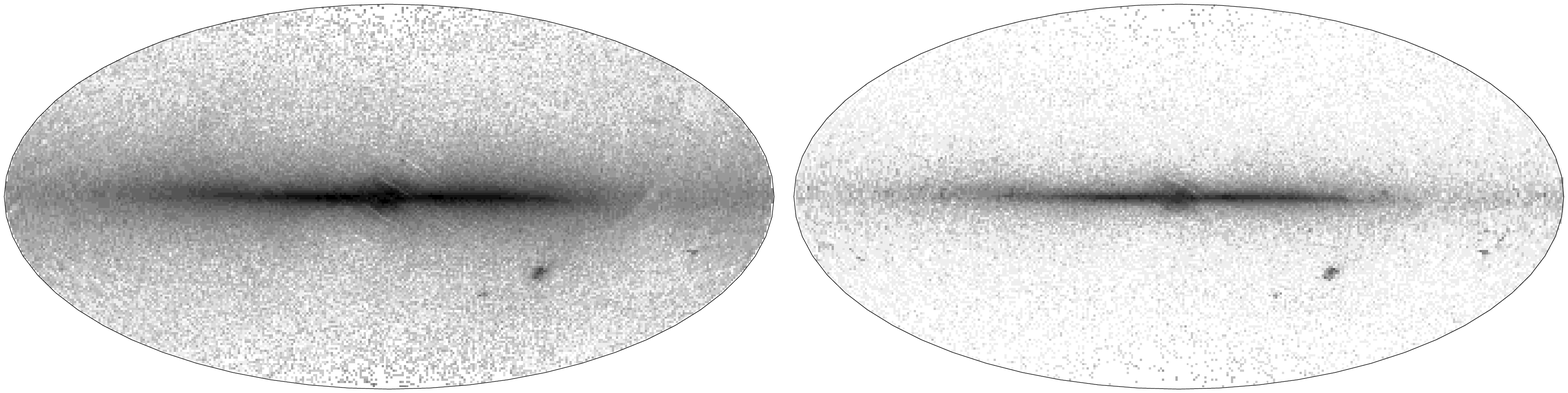}
\caption{Spatial number density distribution
of the detected sources in the Galactic coordinates of the Aitoff projection.
(Left) 9\,$\mu$m sources of 851,189 and
(right) 18\,$\mu$m sources of 195,893.}
\label{fig:psc_allsky}
\end{figure*}

Fig.~\ref{fig:psc_allsky}
shows the spatial distribution of the number density of the catalogued sources
in the 9 and 18\,$\mu$m bands after removal of the rejected events.
One can easily appreciate the presence of highly populated
regions, which correspond to the Galactic plane, the Large Magellanic
Cloud (LMC), the Small Magellanic Cloud (SMC), as well as the nearby
star forming regions such as $\rho$ Oph, Orion and Taurus. Note that
the 18\,$\mu$m sources are strongly concentrated near the thick Galactic
plane, whereas the 9\,$\mu$m source counts fall off smoothly with the galactic
latitude.

\subsection{Flux accuracy \label{flux_acu}}
\begin{figure*}[h]
\includegraphics[width=9cm]{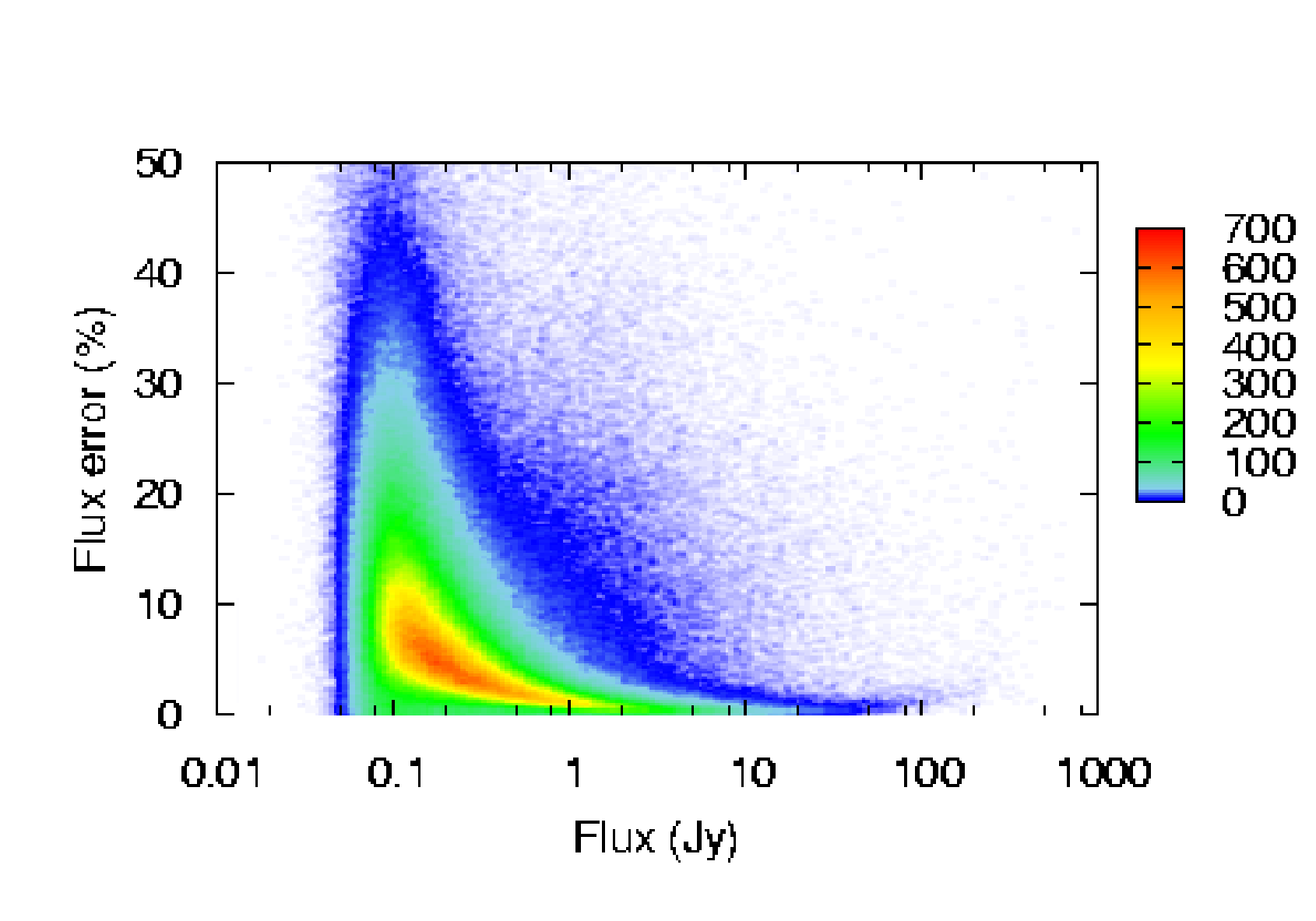}
\,
\includegraphics[width=9cm]{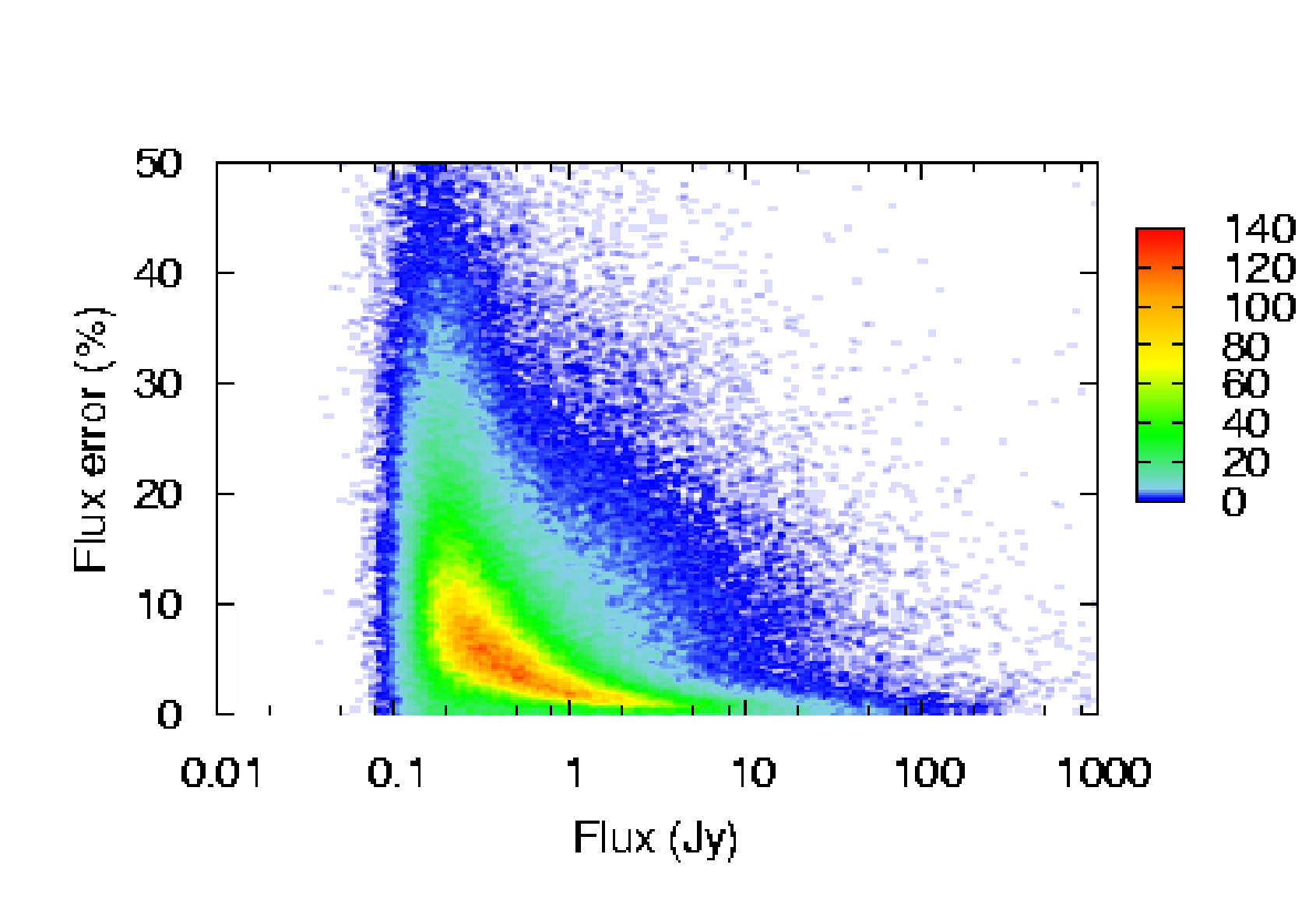}
\caption{Relative flux errors (\%)
for 9\,$\mu$m (left) and 18\,$\mu$m sources (right)
as a function of flux plotted in the density maps.}
\label{fig:psc_fvse}
\end{figure*}

The histogram in 
Fig.~\ref{fig:psc_fvse} shows 
the relative flux errors
in the 9 and 18\,$\mu$m bands as a function of flux.  As expected, the
largest errors are associated with sources with the lowest fluxes.
We find that the most
probable error is 2--3 \% and the
probability of larger errors is small.  The
relative errors are smaller than 15\% for 80\% of the sources and
smaller than 30\% for 96 \% of the sources. It should be stressed
that at this stage we cannot quantify how much variable objects
``artificially'' contribute to an increase of the scatter in the observed
fluxes.
The inclusion of mid-IR variability of the objects will be considered
in future catalogue releases.

\begin{figure}[h]
\center
\includegraphics[width=9cm]{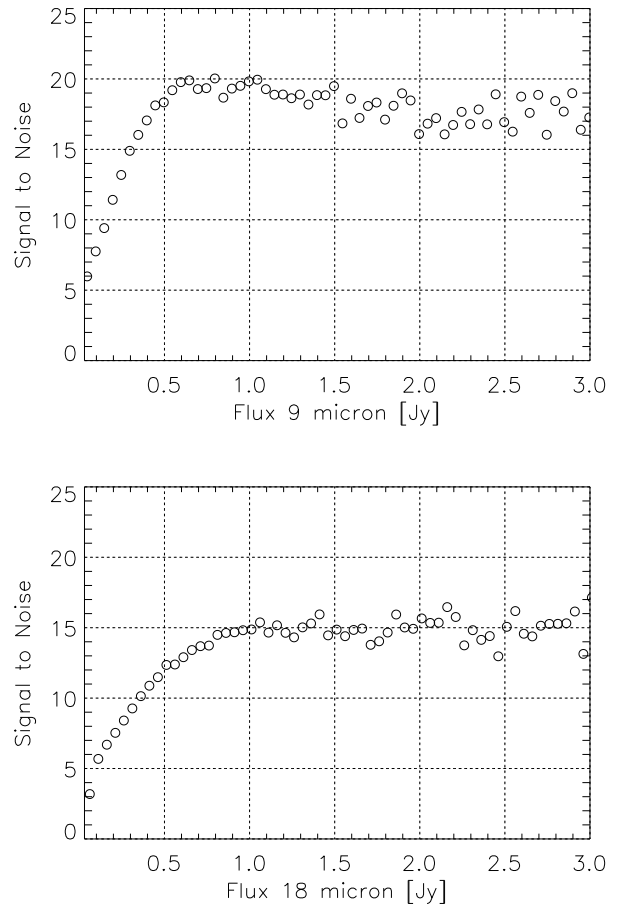}
\caption{Signal to noise ratio as a function of the 9\,$\mu$m
(top) and 18\,$\mu$m band flux (bottom).
}
\label{fig:sn}
\end{figure}

Fig.~\ref{fig:sn} shows
a typical signal-to-noise ratio (S/N)
associated with the different flux levels.
The figure indicates that 
the expected S/N is $\sim$ 6 for the faintest sources of $\sim$ 0.045\,Jy in the 9\,$\mu$m band,
while it is $\sim$ 3 for $\sim$ 0.06 Jy sources in the 18\,$\mu$m band.
The figure also shows that
the S/N increases with increasing flux, 
but
above 0.6\,Jy and 0.9\,Jy it becomes constant or slightly
decreases to $\sim$ 20 and $\sim$ 15
for 9\,$\mu$m band and 18\,$\mu$m band, respectively. 
This leveling-off of the S/N is due to 
uncertainties in the data reduction process.
Errors in the correction for the detector reset anomaly, linearity,
and flat-fielding 
limit accuracy of the  of the flux measurement.

\subsection{Position accuracy}

\begin{figure}[h]
\includegraphics[width=9cm]{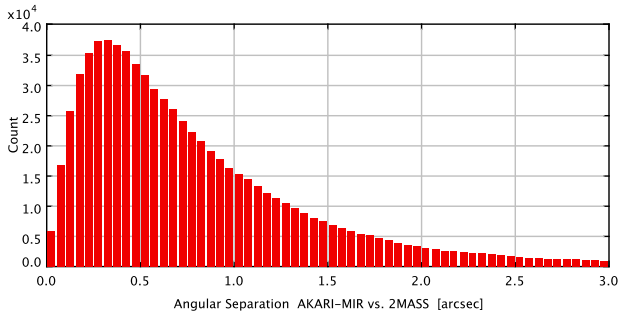}
\caption{Histogram of the angular separation between AKARI coordinates
and the 2MASS coordinates for the common sources.}
\label{fig:casfig3}
\end{figure}

A test of the {\it AKARI} position accuracy is indicated
in Fig.~\ref{fig:casfig3}, which shows the angular separation 
between {\it AKARI} sources and the nearest 2MASS source \citep{2MASS}.
The figure indicates
that about 73\% of the sources have an angular separation $<1''$,
nearly 95\% of the sources have a separation $<2''$.
On average, the mean separation between {\it AKARI} and 2MASS
coordinates of matching sources is 0.8$\pm$0.6$''$.

\subsection{Number of detections per source}

\begin{figure}[h]
\includegraphics[width=9cm]{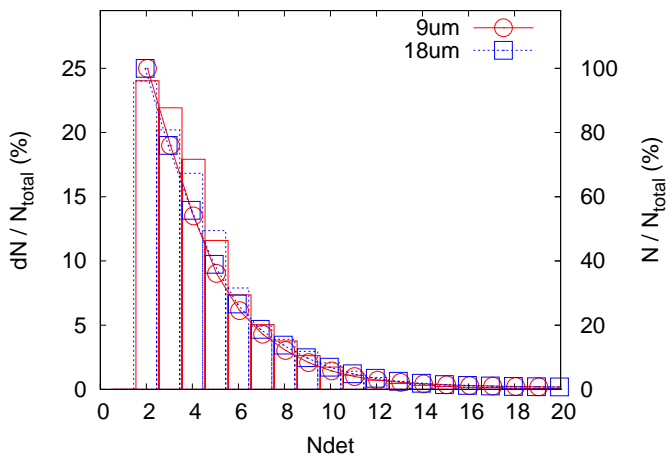}
\caption{Number of detections per source
at 9\,$\mu$m (solid lines) and 18\,$\mu$m source (dotted lines).
The differential counts are shown as histograms in the scale
on the left vertical axis
and the integrated counts are shown by the lines with symbols in the scale on the
right vertical axis.}
\label{fig:ndet}
\end{figure}

The histogram in Fig.~\ref{fig:ndet} shows
the number of detections per source,
both in differential and integrated counts.
As described in \S\ref{cat_compile},
all the catalogued sources are detected at least twice.
Over 80\% of the sources have more than 3 detections and
about half of the sources have more than 5 detections
(right side scale of Fig.~\ref{fig:ndet}).
The average number of detections of the catalogued sources is 6.

\subsection{False detections and missing sources}

\paragraph{False detections}

False detections are in principle rejected
in the two-step source confirmation process.
However, there is an off-chance of including
(1) high energy particle hit events,
(2) slow moving objects,
(3) ghosts produced in the camera optics, and
(4) those triggered by the signal saturation.

\paragraph{Missing bright sources}

\begin{figure}
\includegraphics[width=8.5cm]{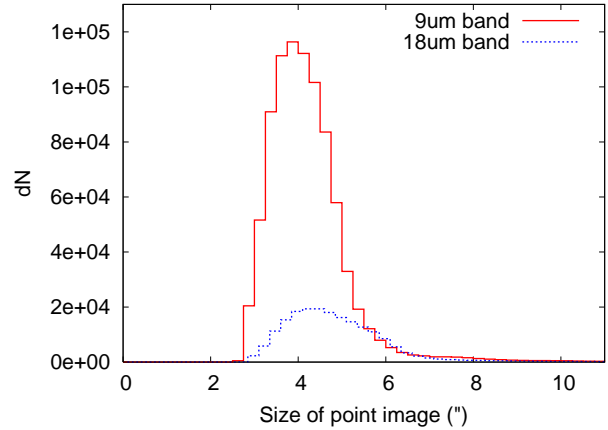}
\caption{Histograms of the spatial extension of {\it AKARI} sources
at 9\,$\mu$m and 18\,$\mu$m, (solid and dashed lines, respectively).
The spatial extension is calculated as $(a+b)/2$, where
$a$ and $b$ are the length in arcsecond for the major axis
and the minor axes of the source image, respectivel.}
\label{fig:extend}
\end{figure}

Some of the bright sources which have {\it IRAS} measurements
are not included in the catalogue because they are
(1) not recognized as a point source with the beam size of {\it AKARI}
because of the spatial extension (Fig.~\ref{fig:extend} shows histograms
of the spatial size of the
{\it AKARI}/MIR sources),
(2) located in the area not covered by this survey,
or
(3) observed only once.
A total 1722 {\it IRAS} sources
with the good quality flag ($f_{\rm qual12}=2$ and $f_{\rm qual25}=3$)
are not included in both the 9\,$\mu$m and 18\,$\mu$m sources.

%

\subsection{Source counts versus flux}

\begin{figure*}
\begin{center}
\includegraphics[width=7cm]{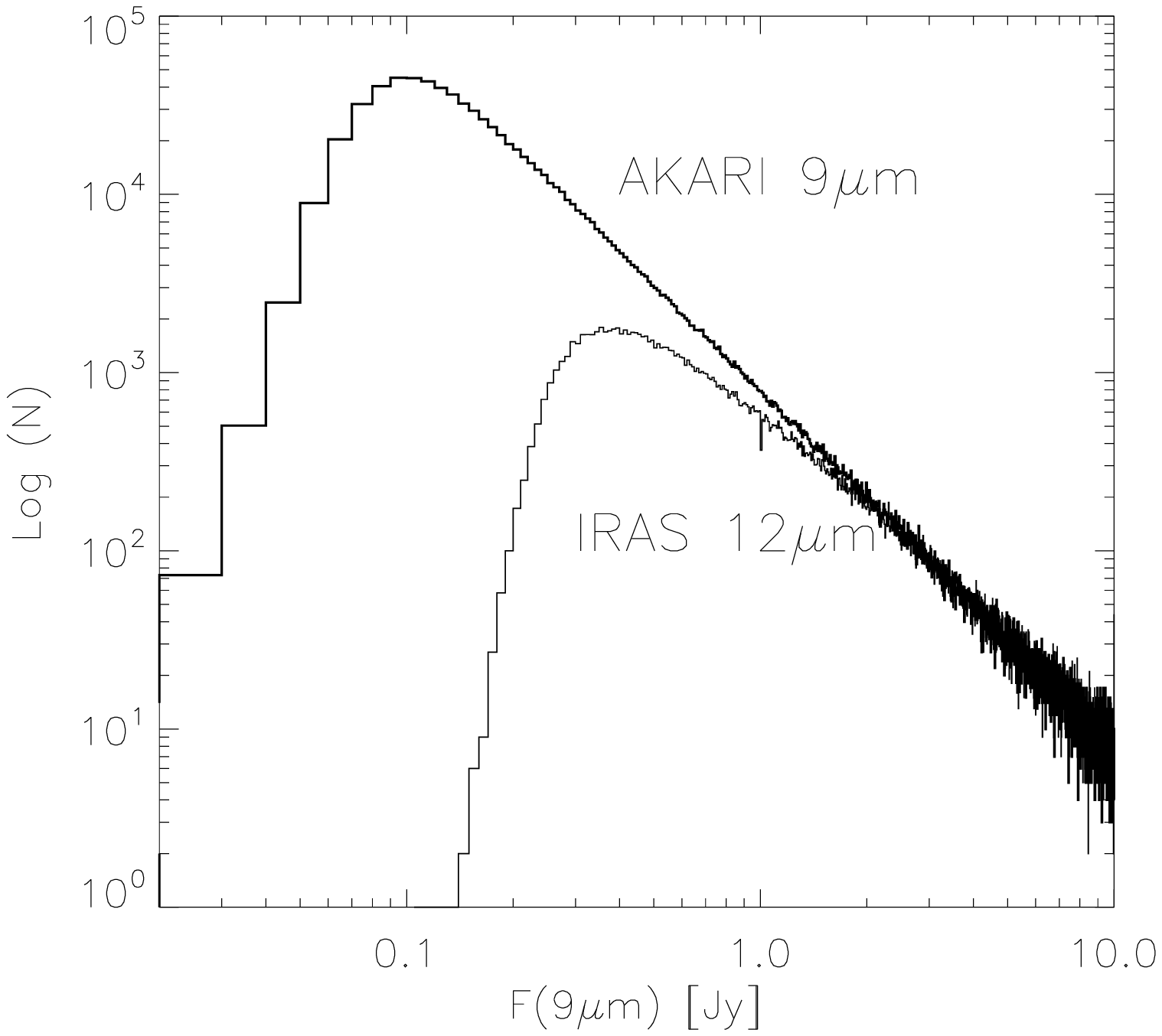}
\includegraphics[width=7cm]{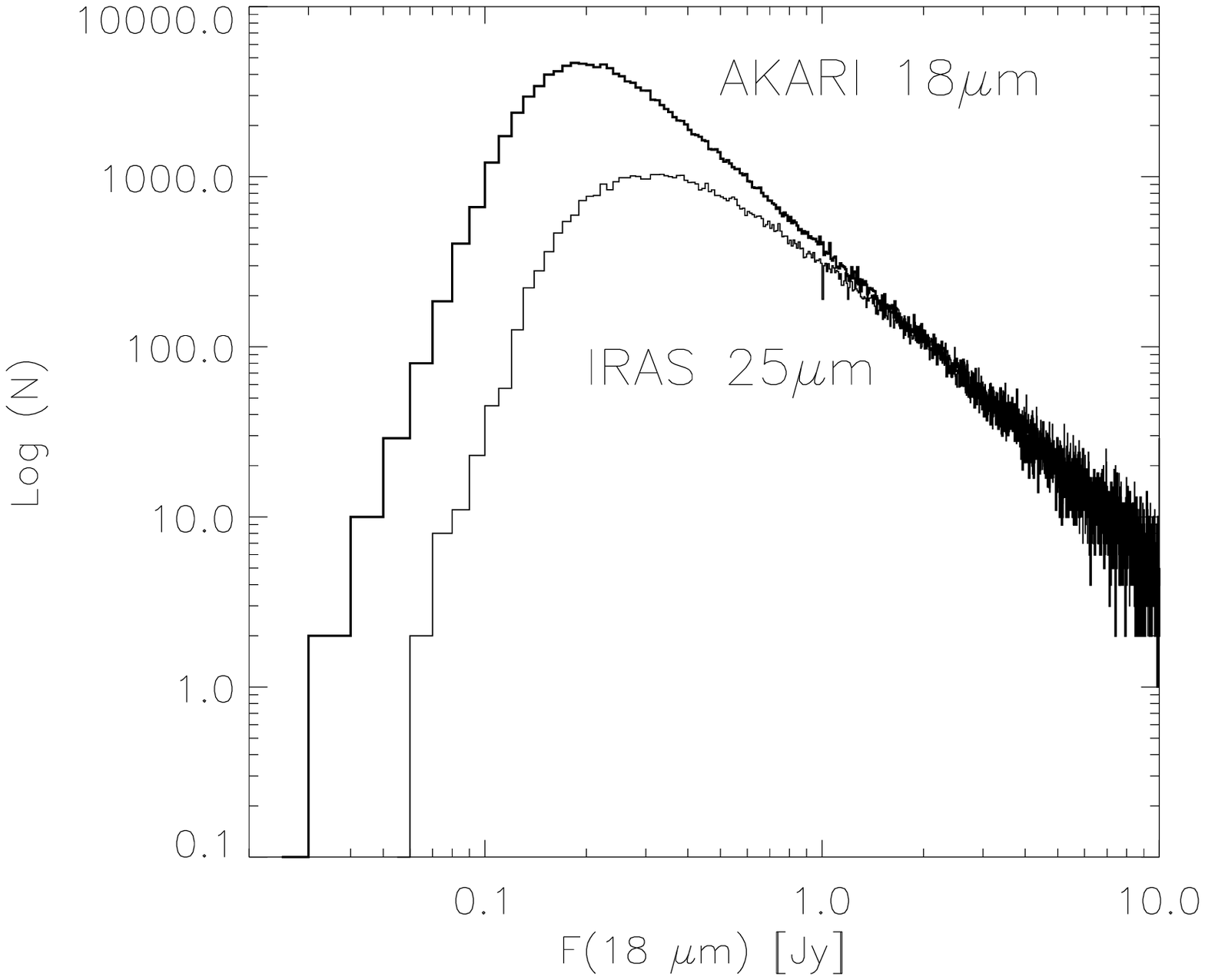}
\caption{The distribution of source counts as a function of the 9\,$\mu$m and
  18\,$\mu$m flux for {\it AKARI} is compared with that from the {\it IRAS}
  survey at 12\,$\mu$m and 25\,$\mu$m.}
\label{fig:akairas}
\end{center}
\end{figure*}

\begin{figure*}
\includegraphics[width=7.5cm]{lfS.epsi}
\,
\includegraphics[width=7.5cm]{lfL.epsi}
\caption{Flux distributions of the detected sources. 
(Left) Flux distribution for the 9\,$\mu$m sources with 18\,$\mu$m detections (dashed),
without 18\,$\mu$m detections (dotted) and total (solid).
(Right) Flux distribution for the 18\,$\mu$m sources with 9\,$\mu$m detections (dashed),
without 9\,$\mu$m detections (dotted) and total (solid).
}
\label{fig:lf}
\end{figure*}

It is instructive to compare the overlall distribution of the
source counts between the {\it AKARI} and IRAS PSCs.  The histograms of the source
counts in the two surveys are compared in Fig.~\ref{fig:akairas}.  The
large difference in terms of the lower cutoff flux in the source counts
between the two surveys is a clear signature of the considerably higher
sensitivity, up to a factor of nearly 10, of the {\it AKARI} survey compared to
IRAS.
  
The distribution  of  source counts  
of the subsets of the 9\,$\mu$m sources with and without a
18\,$\mu$m counterpart are shown
in the left panel of Fig.~\ref{fig:lf}. 
Similarly that for the  18\,$\mu$m sources with and 
without a 9\,$\mu$m counterpart is shown in the right panel of Fig.~\ref{fig:lf}.
Most of the 18\,$\mu$m sources have 9\,$\mu$m detections,
whereas fainter 9\,$\mu$m sources do not have their 18\,$\mu$m counter
parts.
 
\subsection{Completeness}

The completeness of a survey above a given flux level
is usually defined as the fraction of true sources
that can be detected above that level.
It is difficult to apply this concept to the {\it AKARI} survey
because one should dispose of a statistically significant sample
of true sources with known IRC fluxes.
The standard
stars used for the {\it AKARI}/IRC PSC calibration might not represent
a statistically
significant sample in view of the rather poor coverage at
low flux levels.

\begin{figure}[!htpb]
\begin{center}
\includegraphics[width=8cm]{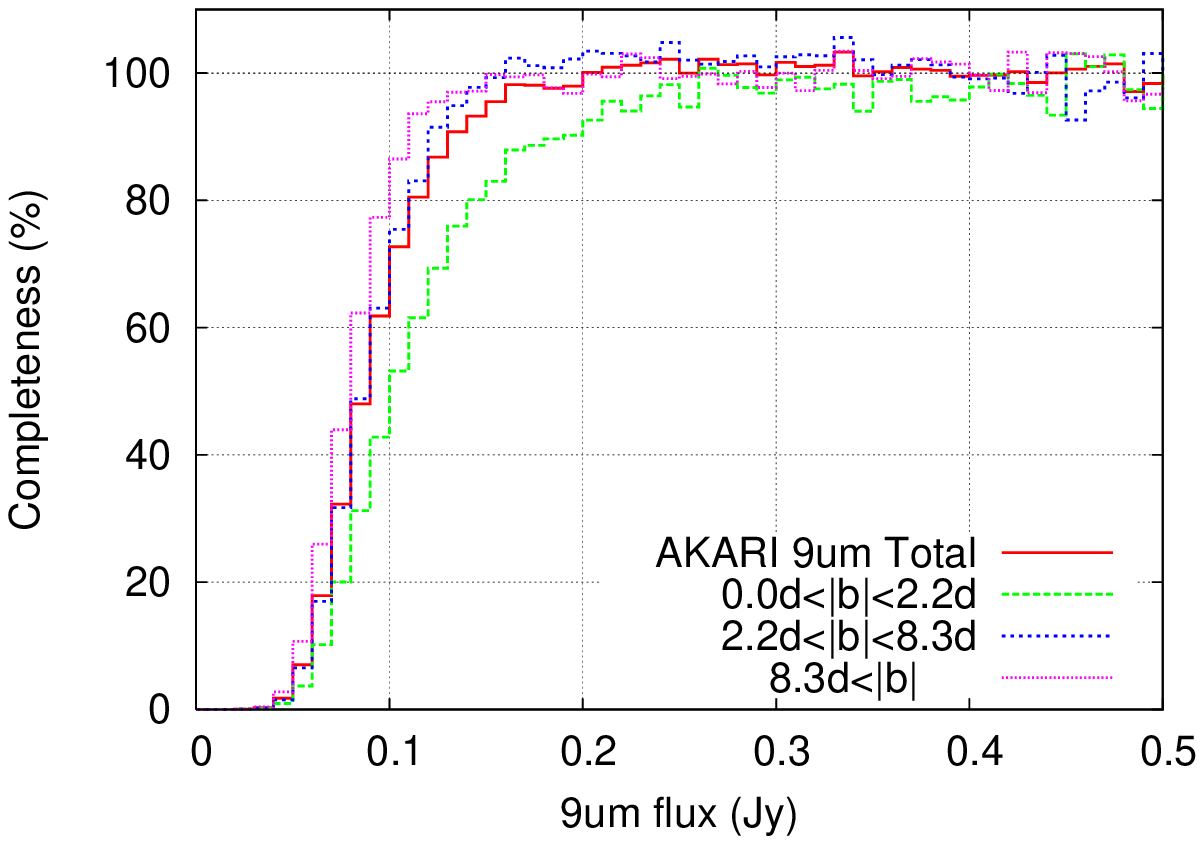}
\includegraphics[width=8cm]{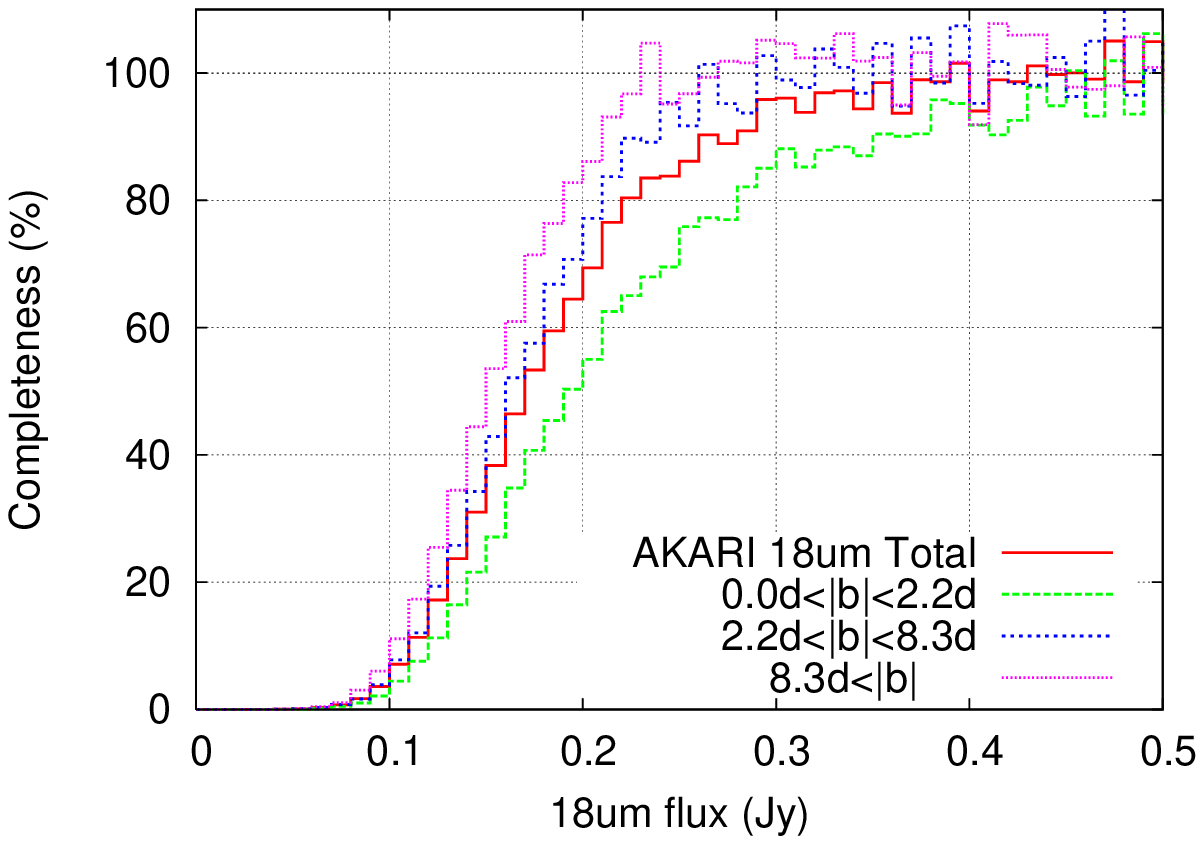}
\caption{
Completeness ratio of the AKARI/IRC survey in the 
9\,$\mu$m and 18\,$\mu$m bands.
Plots are made also for each galactic regions, $|b|<2.2^\circ$,
$2.2^\circ<|b|<8.3^\circ$, and $|b|<8.3^\circ$,
separately.
}
\label{wcomplet}
\end{center}
\end{figure}

To assess the completeness of the {\it AKARI}-MIR survey
we have thus taken a different approach
based on the distribution of sources according to their flux.
Fig.~\ref{fig:akairas} 
shows an interesting feature:
source counts decline exponentially with increasing flux after a peak value
is reached around 0.1 Jy (9\,$\mu$m band) and 0.2 Jy (18\,$\mu$m band).
One can thus make a reasonable assumption that the exponential decay is an
intrinsic property of source counts at the relevant wavelengths and,
on this basis, one can define completeness as the ratio of the number
of sources actually observed by the number of sources predicted by the
above equations. The results of completeness
are shown in Figure \ref{wcomplet}. From this figure we deduce
the completeness ratios reported in Table \ref{castab6}.

To evaluate the detection limit of the survey,
we can make use of the S/N characteristics
shown in Fig.~\ref{fig:sn}.
>From this figure one can deduce that,
for S/N $\sim$ 5, the detection limit is
about 0.05 and 0.09 Jy in the 9\,$\mu$m and 18\,$\mu$m bands, respectively.

A summary of the completeness of the survey at various flux levels, together
with the corresponding values of the S/N ratio are given in 
Table \ref{castab6}.
These results are in fairly good agreement with a
pre-launch prediction in Ishihara et al. (2006a).

The completeness is also tested separately
for different Galactic latitude ranges,
$|b|<2.2^\circ$, $2.2^\circ<|b|<8.3^\circ$, and $|b|>8.3^\circ$,
where the division is made to have roughly similar numbers of 
the objects.
Fig.~\ref{wcomplet} indicates that the completeness becomes
worse in the Galactic plane ($|b|<2.2$ degree),
which can be attributed to either
the source confusion or the spatially variable background.

\begin{table}[!htpb]
\caption{Completeness and signal-to-noise ratio}
\label{castab6}
\begin{center}
\begin{tabular}{lrrrr} \hline\hline
\multicolumn{5}{c}{9\,$\mu$m band} \\ \hline
Completeness & 5\% & 50\% & 80\% & 100\% \\
Flux[Jy] & 0.07 & 0.10 & 0.11 & 0.15 \\
S/N      & 6.7  & 7.6  & 8.5 & 9.8 \\ \hline
\multicolumn{5}{c}{18\,$\mu$m band} \\ \hline
Completeness & 5\% & 50\% & 80\% & 100\% \\
Flux[Jy] & 0.10 & 0.16 & 0.21 & 0.28 \\
S/N      & 5.0  & 6.5  & 7.6 & 8.7 \\ \hline
\end{tabular}
\end{center}
\end{table}

\subsection{Comparison with the IRAS catalogue}

To further evaluate the reliability of the {\it AKARI} flux
  measurements at 9\,$\mu$m and 18\,$\mu$m, we have carried out a comparison
  of these with the {\it IRAS} 12\,$\mu$m and 25\,$\mu$m fluxes for common
  sources.  

\begin{figure*}
\begin{center}
\includegraphics[width=6.5cm]{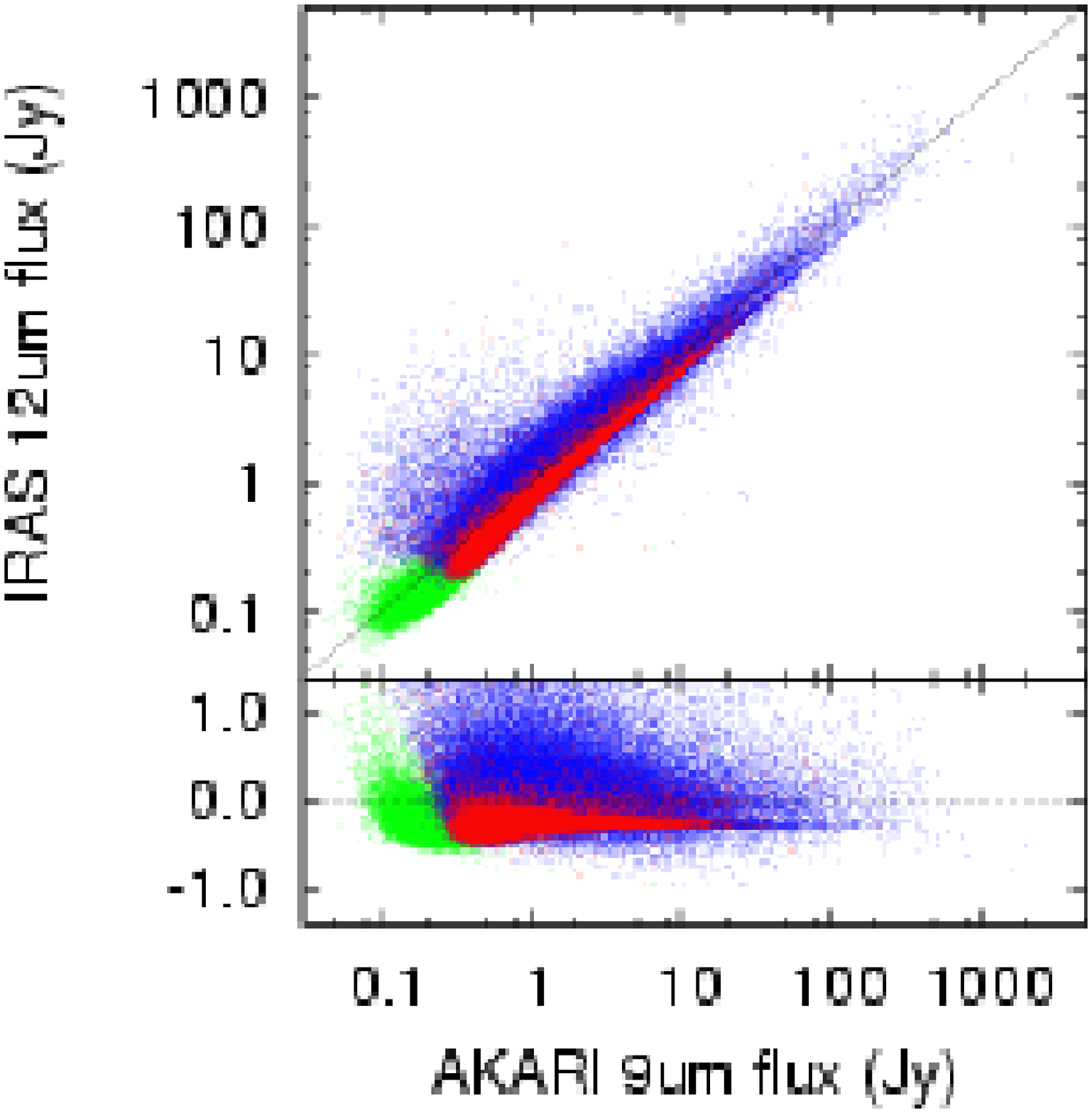}
\includegraphics[width=6.5cm]{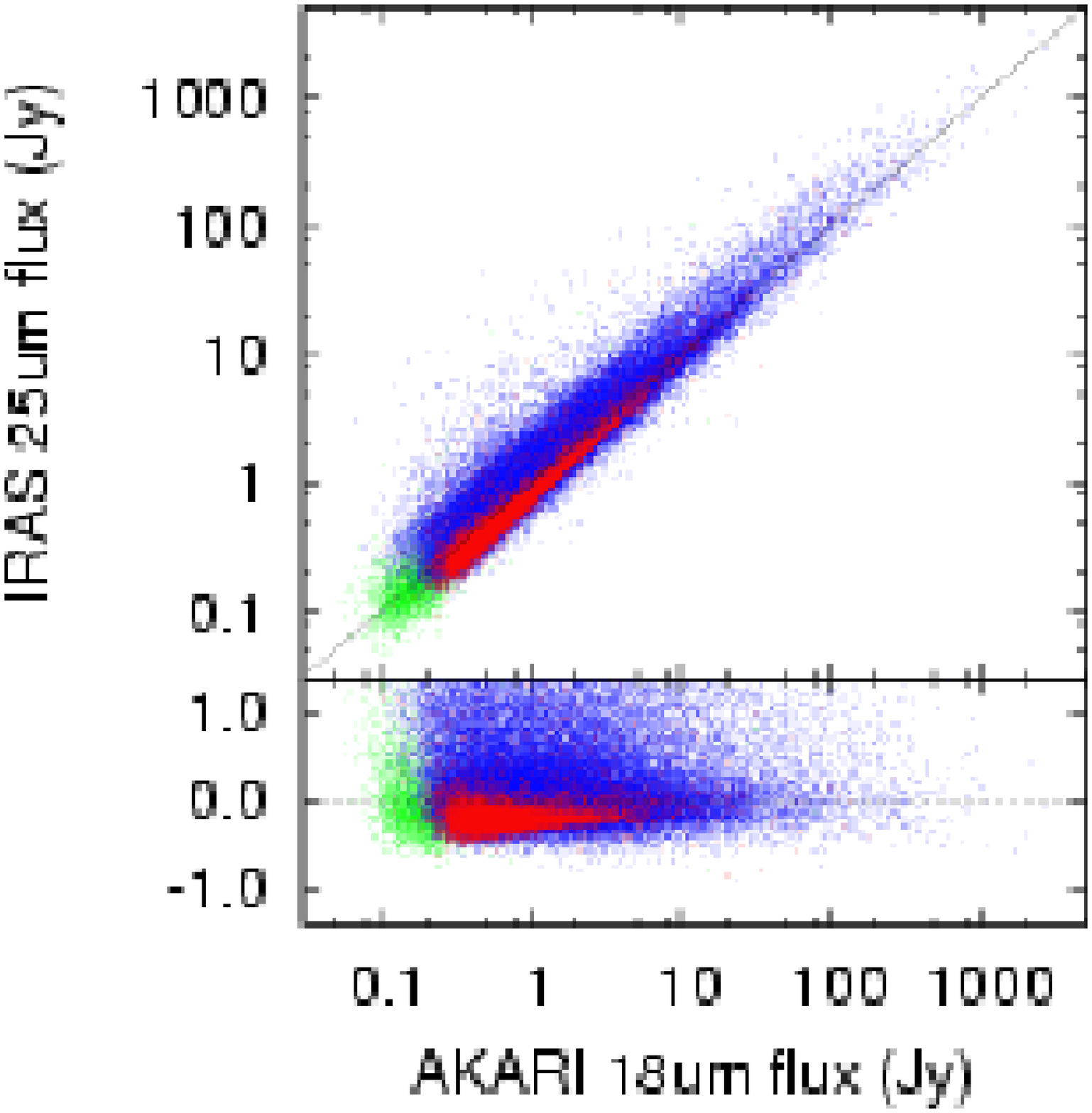}
\caption{(Left) Comparison of the {\it AKARI} 9\,$\mu$m versus IRAS 12\,$\mu$m 
fluxes for the cross-identified sources.
The red dots indicate sources with $f_{\rm qual12}=3$ in the IRAS PSC
labeled as a star in the SIMBAD database.
The blue dots shows the other sources with $f_{\rm qual12}$=3 in the IRAS PSC.
The green dots indicate IRAS FSC sources.
(Right) Comparison of the {\it AKARI} 18\,$\mu$m versus IRAS 25\,$\mu$m fluxes
for the cross-identified sources.  The symbols are the same as in the left
panel.}
\label{fig:cmp_iras}
\end{center}
\end{figure*}

In Fig.~\ref{fig:cmp_iras} a comparison is given between {\it
    AKARI} and {\it IRAS} fluxes observations for the cross-identified
  objects. It clearly appears that there is a close correlation between the
  two sets of data. This is especially true for stellar sources; indeed, many MIR
  sources such as compact \ion{H}{ii} regions, reflection nebulae and
  planetary nebulae, have extended emission or are located within extended
  objects.  In such cases, IRAS measures the total flux of the extended
  emission because of the larger apertures
($0\farcm75\times4\farcm5-4\farcm6$ pixel size), whereas
  {\it AKARI} measures the flux of the peak emission on the extended objects
  with a smaller aperture ($\sim9\arcsec$ beam size) as demonstrated in the
  case of a reflection nebula \citep{IC4954}.  Some sources are
  brighter in the IRAS PSC than in the {\it AKARI} PSC.  The difference could be
  attributed to the effect of the difference in the spatial resolution.

%
%

\section{Summary}
The {\it AKARI} Mid-Infrared all-sky survey was performed
with two mid-infrared broad bands centered at 9 and 18\,$\mu$m. 
More than 90\% of the entire sky was observed in both bands.
A total of 877,091 sources
(851,189 for 9\,$\mu$m, 
195,893 for 18\,$\mu$m) are detected
and included in the present release of the point source catalogue.
This {\it AKARI} Mid-Infrared point source catalogue 
is scheduled for the public release in 2010
after the prioritized period for the team members \citep{relnote}.
We present the
spatial distribution, flux distribution, 
flux accuracy, position accuracy, and completeness of the 
sources in the {\it AKARI} MIR All-Sky Survey Catalogue version $\beta$-1.

The {\it AKARI} mid-infrared survey
provides a unique data-set relevant to
interstellar, circumstellar-, and extra galactic astronomy.
This new deep, large survey
is well suited to research in the various fields of astronomy,
such as search for warm debris disks,
of asteroid analogue (Fujiwara et al. 2009a, 2009b),
and provides valuable information
for the study of planet formation and other fields.
The chemical compositions of an unprecedented number of dust forming 
asymptotic giant branch (AGB) stars can also be investigated \citep{Ita}
making use of the characteristics of the filter bands of this survey.
Such studies enable new discussions on the structure, the star formation history and
the cycle of matter in our Galaxy.
The mid-infrared all-sky survey also provides a chance to detect
highly obscured active galactic nuclei (AGNs),
which are difficult to observe in
previous optical or X-ray surveys.
The population of obscured AGNs is expected to contribute
to the hard X-ray background, which is not
competely resolved into individual sources yet \cite{Ueda}.

\begin{acknowledgements}
This research is based on observations with {\it AKARI},
a JAXA project with the participation of ESA.
This research has made use of the SIMBAD database,
operated at CDS, Strasbourg, France.
This work was supported by a Grant-in-Aid for Scientific
Research on Priority Areas from the Ministry of Education, Culture,
Sports, Science and Technology, Japan (No. 16077201).
H.F. is financially supported by the Japan Society for
the Promotion of Science.
S.H. was supported by Space Plasma Laboratory, ISAS, JAXA.

\end{acknowledgements}

\appendix

\section{Nest 4$\times$4 mode operation and unit image construction\label{nest4x4}}

\begin{figure*}[h]
\center
\includegraphics[width=14cm]{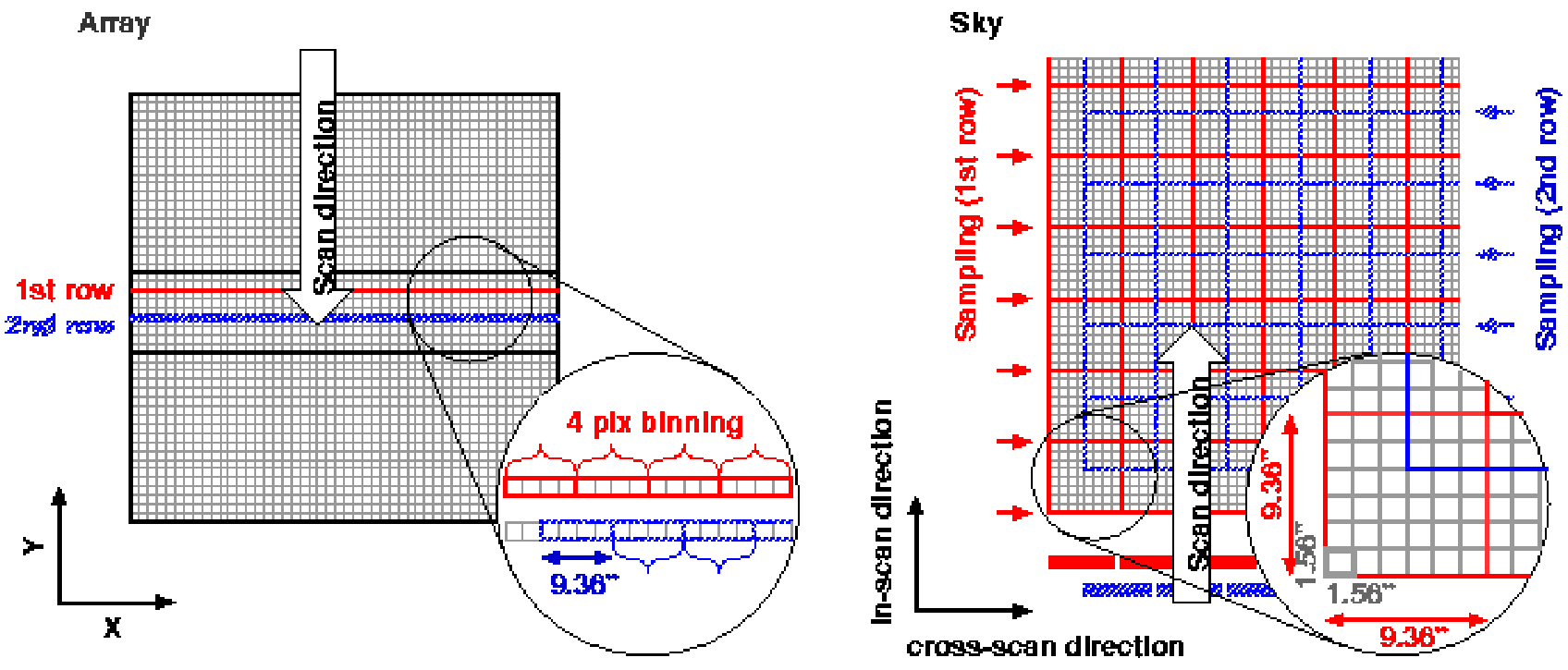}
\caption{(Left) Illustration of the nest 4$\times$4 mode
array operation in the survey mode.
(Right) Image reconstruction of the data
taken in the nest 4$\times$4 mode operation.}
\label{fig:4x4}
\end{figure*}

\begin{figure*}
\center
\includegraphics[width=10cm]{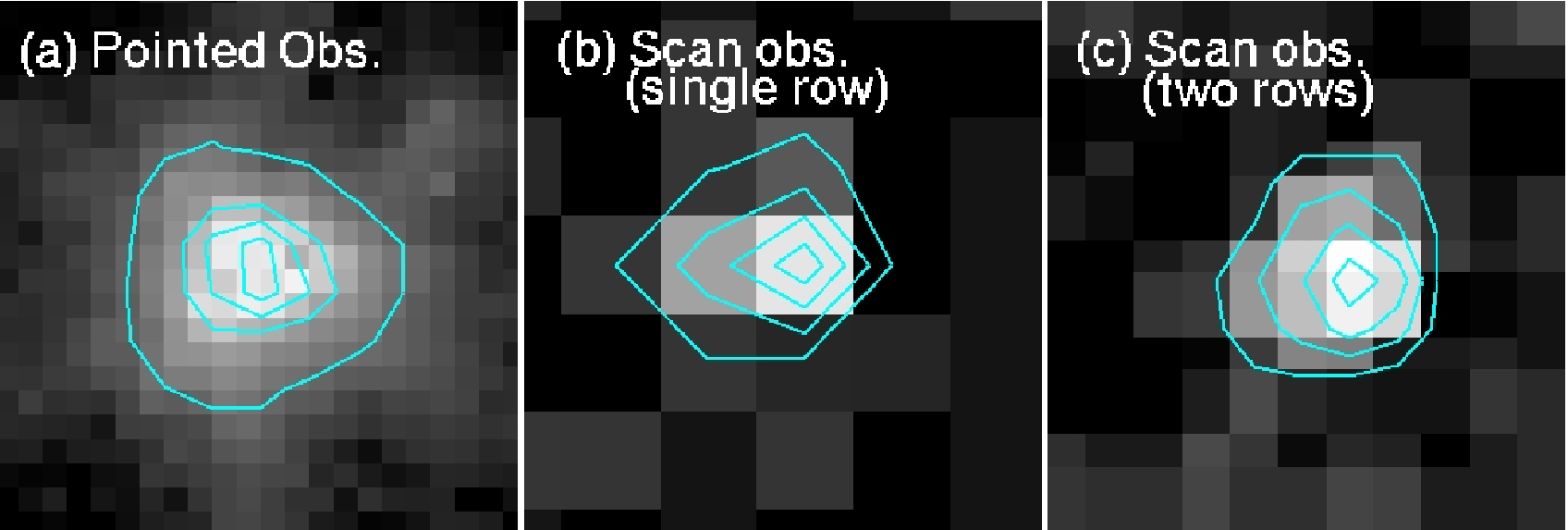}
\caption{9\,$\mu$m image of the standard star (HD 42525):
(a) the full-resolution image obtained in the pointed observation,
(b) the images processed from the single-row data in the survey observation,
and 
(c) the images reconstructed from the two-row data in the survey observation.
}
\label{fig:psf}
\end{figure*}

Here we expand on the detail
of the data acquisition in the all-sky survey mode and
subsequent image reconstruction method
to describe the pixel scale (Fig.~\ref{fig:4x4}).
Though
the original pixel scale of the MIR-S camera is $2\farcs34\times2\farcs34$
($2\farcs51\times2\farcs39$ for MIR-L),
we degrade the spatial resolution in the survey observation
to reduce the amount of output data
to meet the downlink capacity.
%
The pixel scale of the data taken by a single row
in the in-scan direction is adjusted as $9\farcs36$
by setting the sampling rate as 22.7\,Hz
under the scan rate of the satellite of 216$''$ s$^{-1}$.
The pixel scale in the cross-scan direction is adjusted as $9\farcs36$ by 
binning of the output of four adjacent pixels.
The resulting effective pixel scale
obtained in the scan by a single row 
(hereafter virtual pixel scale) is 4 times larger than
the original pixel scale. 
%
The scan observation was made by two rows
to enable two independent observations
in the milli-seconds interval.
The sampling timing and the combination of the binning pixels 
are adjusted
to construct two independent grids on the sky.
Because the scan rate was changed from the value expected before launch
due to the change in the altitude from the planned 750\,km to 700\,km,
the shift of the grids in the in-scan direction
is not exactly a half of the virtual pixel size.

Fig.~\ref{fig:psf} compares an image of the same star
constructed from a single-row observation in the survey mode,
that processed from a two-row observation in the survey mode,
and that obtained in the imaging mode in the pointed observation
with the original pixel scale.
The FWHM of the PSF for the 9\,$\mu$m and 18\,$\mu$m bands
is $5\farcs5$ and $5\farcs7$, respectively.
The virtual pixel size is larger than the PSF size.
However the process of the two-row observations in the survey mode
reconstructs the image with the spatial resolution
compatible with that obtained in the pointed observations
with the original pixel size (center of Fig.~\ref{fig:psf}).

Thus the process of the two-row observation allows us to make
(1) confirmation of the source detection in milli-seconds after,
(2) reduction in the output data rate,
and (3) higher spatial resolution than that in the single-row operation.
Apparent asymmetry seen in the pointed data (Fig.~\ref{fig:psf} (a))
can be attributed to the effect of the telescope truss.
Part of the asymmetry seen in the survey data may also be
attributed to the undersampling operation.
No variation of the PSFs is recognized in the 9\,$\mu$m and 18\,$\mu$m bands during the
liquid Helium period in the pointing mode data.

\end{document}